# Equilibrium Configurations of Cantilever under Terminal Loads

Milan Batista

University of Ljubljana, Faculty of Maritime Studies and Transport,

Pot pomorščakov 4, Portorož, Slovenia, EU

**Abstract**

The paper provides an exact analytical solution for equilibrium configurations of a cantilever rod subject to inclined force and torque acting on its free end. The solution is given in terms of Jacobi's elliptical functions and illustrated by several numerical examples and several graphical presentations of shapes of deformed cantilever. Possible forms of a cantilever's underlying elastica are discussed in detail and various simple formulas are given for the calculation of characteristic dimensions of elastica. For the case when a cantilever is subject only to applied force four load conditions are discussed: follower load problem, load determination problem, conservative load problem and rotational load problem. For all the cases the formulas or effective procedure for the solution is provided.

**1 Introduction**

In this article we discuss the problem of the determination of the deflection curve of an in-plane elastic cantilever rod subject to various forms of terminal force. The origin of the problem traces back to Galileo, who in 1638 posed two problems concerning the construction of a cantilever. Over the following decades these problems, through the works of Hook, Mariotte and Leibniz, gradually yield to the question of determining the deflection curve of a cantilever. By 1691 the problem was narrowed by James Bernoulli to the special case when terminal weight is acting on a column; this was finally solved in the general case by Leonhard Euler in his famous treatise on elastic curves in 1743 [1] after a period of correspondence with Daniel Bernoulli. In his treatise, Euler by parametric study of the solution enumerated nine possible equilibrium shapes for the infinite rod under equal but oppositely directed forces and then applied the classification to a cantilever. He found that a cantilever can only be bent to six shapes, taking into account only the non-inflection parts of an underlying elastic curve. The solution is presented in the form of two non-elementary integrals using their power series expansions to make practical calculations. He also provided the formula for what we now call the critical force.

Upon the development of the theory of elliptic integrals and elliptic functions in the 19th century, researchers sought to obtain a closed form solution of the problem. One such solution was given by Clebsh [2] (§53, pp 218-222), who considered a column under a vertical force but did not refer to elliptic integrals. In 1880, Saalschütz [3] published a treatise that was entirely devoted to the determination of deflection curves of a cantilever under inclined force by using Legendre's elliptic integrals of the first and second kind. In this book we can find the closed form expressions for determining the shape of a deflected cantilever, and special expressions for the displacement of its free end, when subjected to either inclined, transversal or axial force. The closed form solution in terms of Jacobi's elliptic functions was given in 1885 by Hess [4](Eq 18-19), who studied rods using Kirchhoff's kinetic analogy, which states that the equations pertaining to the elastic rod are equivalent to equations describing the motion of the rigid pendulum. Hess used Jacobi's notation of elliptic functions. The solution in Gudermann's notation was provided in 1893 by Love [5] ( §228 pp 49-54). Both these solutions are, however, for the case of a rod under two oppositely directed forces and are not directly applicable to a cantilever. We note that Love called elastic curves with inflection points (corresponding to an oscillating pendulum) inflectional and those without inflection points (corresponding to a revolving pendulum) non-inflectional. Later editions of Love's book use his shortened version of the section about elastic lines. Max Born [6] in his dissertation from 1906 conducted the first experimental theoretical study of the post-buckling equilibrium configurations of a cantilever by using an elliptic integral solution. (For more historical data about the rod problems we refer to Todhunter [7], Timoshenko [8], Truesdell [9], Goss [10] and Levien [11].)

In the first half of the 20th century numerous authors used or rediscovered Legendre's elliptic integrals form for solving the cantilever problem. Malkin [12] discussed large deformations of elastic columns under terminal weight. Hummel and Morton [13] used the solution to implicitly measure Young's modulus of the cantilever rod. Barten [14] provided an expression for the vertical deflection of the free end point of a cantilever loaded by transversal force, while Bisshopp and Drucker [15], considering the same problem, also derived an expression for its free end axial displacement. Expressions for transversal and axial displacement of an axially loaded column can also be found in Timoshenko [16] (pp 76-82). Yet another derivation of an elliptic integral solution for the deflection of a cantilever under inclined force - using somewhat extensive notation – was given by Mitchell [17].

Until the appearance of digital computers, the calculations of cantilever deflection were made by using tables of elliptic integrals. Various approximate methods were proposed to overcome this difficulty. Beth and Wells [18] provided a power series solution of the problem for an inclined force that is applicable for moderate cantilever deflection. Another power series solution for a transversally loaded cantilever, which results from a variant of the successive approximation method, was obtained by Scott et al. [19]. For inclined force, Frisch-Fay [20, 21] suggested a method by which a cantilever is broken into segments that are identical to a vertically loaded column and in this way replaced integration with the solution of transcendental equations resulting from the condition of a smooth connection between the successive cantilever segments. The same author also published a valuable book treating flexible rods [21] in which a chapter is devoted to the cantilever problem. Massoud [22] considered a cantilever under transversal force and provided approximate formulas of deflection of the free end, derived by the choosing of an axis with a slope that is the average value of the cantilever tangent angle. For future references for the period up to the 1970s to we refer the reader to Schmidt and DaDeppo [23].

The appearance of mainframe computers in the 1960s and the 1970s allowed the use of various numerical techniques for solving the problems relating to the cantilever. For this reason, the problem became the subject of many master's and doctoral theses and therefore beyond this period a future examination of relevant literature in a strict chronological manner is virtually impossible. We thus omit a review of the articles that are closely related to the development of the finite element method (FEM) and the cantilever problem is used as a test example.

In 1981 Wang [24] discussed the problem of deflection of an inclined cantilever subject to a vertical load. For a small and large value of applied force, and for a nearly vertical cantilever under arbitrary valued force he derived an approximate analytical expression using the perturbation method. For the general case he used a numerical method. When one uses numerical methods, technically speaking the cantilever problem is a two point boundary problem (BVP) where one end has fixed geometric conditions and the other end has a prescribed torque. Therefore Wang and later other authors proposed a method that transforms BVP into the initial value problem (IVP) that can be solved by direct numerical integration. Thus Wang suggested a two step method where in each step an initial value problem is solved by the Runge-Kutta numerical integration. In the first step, he, by selecting the value of the cantilever's free end slope, calculated load parameter, cantilever inclination and bending moment at its clamped end. With these data he then, in the next step, computed cantilever deflection. Although Wang pointed out that his numerical method "is much easier than elliptic functions, which also require numerical evaluation," his method does not work if the initial data are load parameter and cantilever inclination at the clamped end. Moreover, with the appearance of low cost computers in the 1980s and the parallel development of numerical algorithms for the calculation of elliptic functions ([25]), the elliptic integral solution became attractive for many researchers for various problems. Thus Mattiason [26] published an article in which he provided tabular values of the displacement and the slant of a transversally loaded cantilever free end as a function of load parameter that can be used as a check of the results of numerical solutions against an exact solution. Lau [27] provided closed form solutions of a cantilever subject to an inclined force and free end torque in the form of elliptic integrals. The same load conditions were considered by De Bona and Zelenika [28] in their article devoted to studying the limits of applicability of elliptic integral solutions in regard to required degrees of accuracy of calculations. Howell and Midha [29] and Saxena and Kramer [30] used the elliptic integral solution as part of a study of large deflections in compliant mechanisms where the latter authors also included the free end bending moment among a cantilever load. Recently Yau [31] consider a guyed cantilever column pulled by an inclined cable (the problem already discussed by Saalschütz [3] (§15)) where he also used the elliptic integral solution. In the 1980s the valuable book of Popov [32] offered extensive analysis of elastic rods by using elliptic integrals.

In 1992 Navaee and Elling [33] published their famous article which dealt with a method for obtaining all possible equilibrium configurations of a cantilever beam under an inclined force. Their starting point was the well-known expression that results from the condition that a cantilever is inextensible, and gives the load parameter as a function of the end slope in the form of a definite integral; i.e., the difference of two incomplete elliptic integrals of the first kind. They observe that the upper and lower limits of the integral can have multiple values, hence, for a given load parameter, it has multiple solutions for the end slope, or, in other words, these multiple solutions yield multiple possible equilibrium forms of cantilever. Once they numerically calculate the value of the end slope they determine coordinates of a deformed cantilever by using the elliptic integral solution. They also consider the question of the number of possible equilibrium configurations, but provide no general conclusion other than that it depends on the value of the load parameter and that the number can be odd or even. A drawback of their discussion is the lack of generality, as they enumerate only seven possible equilibrium configurations and consequently their graph illustrating the distribution of the load parameter versus the end slope is incomplete in that it fails to show that there are an infinite number of branches. The solution that they gave is thus applicable only to values of load parameter up to 12. In their next article the authors establish the possible range of end slope for a given force inclination [34]. A numerical procedure that allows the determination of all equilibrium shapes of cantilever subject to inclined force was later provided by the present author [35].

Until the beginning of the 1980s, researchers mainly consider the conservative load problem. It seems that up to that time the non-conservative problem was considered only through the question of stability of axially loaded columns ([36, 37]). The solution for the cantilever subject to non-conservative transversal force was given by Argyrs and Symeondis ([38]) and Alliney and Tralli ([39]) using FEM, and later Saje and Srpčič ([40]), (which considers extensible beams) using the finite difference method. Recently Shvartsman [41] considered a cantilever subject to inclined force and solved it by using a method similar to Wang's [24] by reducing the problem to an initial value problem and then solving equations by numerical integration. A similar method that transformed BVP to IVP by reverse sense of integration was also used by Nallathambi et al. [42]. The same problem was treated by Mutyalarao et al. [43] by using a semi-analytical approach. As input data they take the value of the cantilever free end slope by which they then calculate the load parameter, expressed as an elliptic integral. With these data the problem becomes an IVP that they then solved using the Runge-Kutta numerical integration.

In 1986 Nageswara and Venkateswara ([44],[45]), in two articles, considered a variant of non-conservative force, which posits that the direction of applied force rotates in relation with the rotation of a cantilever's free end. In the first article they convert the initial BVP into IVP and use the Runge-Kutta numerical integration. In the second article they solve the problem analytically by using an elliptic integral solution. Later Mutyalarao et al. [43] solved the problem in the same manner as described above. We note that all that solutions use as input data cantilever free end slope and then calculate the load parameter.

Several articles published by a group of Russian researchers at the beginning of this century deserve special attention. Zakharov and Zakharenko [46] consider the dynamic instability of a cantilever under a transversal force, viewing it as an eigenvalue problem where the characteristic equation was obtained from the condition that at the free end the bending moment vanishes and expressed their solution by using Jacobi's elliptic functions. For each eigenvalue there is a characteristic critical force and consequently this yields a different number of deflected cantilever inflection points. For them cantilever deflection without inflection points are static and with inflection points dynamic. A similar solution for an inclined force was given by Zakharov and Okhotkin [47] and for a non-conservative inclined force by Zakharov et al [48]. Levyakov [49] and Levyakov and Kuznetsov [50] examine the stability of post-buckling equilibrium states of rods (among others also cantilever) and for the discussion also use Jacobi elliptic function solution.

Recently, some semi-analytical methods were proposed for solving the cantilever problem. Wang et al. [51] provided a solution for the case of a transverse conservative force using the homotopy method, which expresses an explicit approximate solution of the problem in the form of a truncated arc-length parameter power series wherein the series coefficients are calculated numerically. Using the same method Kimiaeifar et al. [52] offered the solution of the problem for a non-conservative inclined force and bending moment [52]. We note that the displayed deflected cantilevers they present do not include inflection points. Tari [53] solved the problem by what he calls the automatic Taylor expansion technique. In essence, he approximates the solution by expanding unknowns as power series' of arc-length parameter. He presents his solutions in graphical form but again none of the displayed deflected cantilevers includes inflection points.

We note that there are numerous articles that deal with the cantilever subject to more complex load and possibly include geometric and/or material nonlinearities. However, since they are not directly related to the present problem, they were not considered.

The aim of this paper is to give yet another analytical solution of cantilever problem where we treat it's possible load conditions from the single point of view. From the review, we see that that there are in essence three analytical approaches to the problem: using Legandre elliptic integrals where independent variable is cantilever tangent angle, using Jacobi elliptical functions where independent parameter is cantilever arc-length and various series expansions. Definitely, the first two methods are superior since they allow to obtained closed form solution that include all possible cantilever equilibrium configurations. In our opinion, among them, the Jacobi elliptical functions are more flexible for a discussion of the problem as are elliptic integrals. Therefore, we will for the solution of the problem use Jacobi's elliptic functions.

The organization of the article is the following. First, we give derivation of basic equations where we, apart for slightly changed notation, follows Antman [54] (Chapter IV). The next two sections are devoted to the solution of basic equations while the fifth section gives some numerical values and some comparison with results of other authors. In the sixth section we in detail discuss possible shapes of cantilever underlying elastica and in seventh section we apply the solution to discuss various force load conditions. The article ends with summary of obtained results.

**2 Formulation of the problem**

*Geometry and equilibrium.* We consider an initially straight inextensible and unshearable elastic rod of length *L* with one end clamped and a force and a torque acting at the other end. In the Cartesian coordinate system *Oxy* the shape of the deformed base curve of the cantilever is described by the following differential equations ([54],pp 87-88)

$$\frac{dx}{ds}=-\cos\phi \qquad \frac{dy}{ds}=-\sin\phi \tag{1}$$

$$\frac{d\phi}{ds} = -\kappa \tag{2}$$

where $x(s)$ and $y(s)$ are coordinates of the base curve, $\phi(s)$ is the angle between tangent to the base curve and the x-axis, $\kappa(s)$ is the base curve curvature and $s \in [0,L]$ is the arc length parameter measured from the cantilever free end to the cantilever clamped end (see Figure 1).

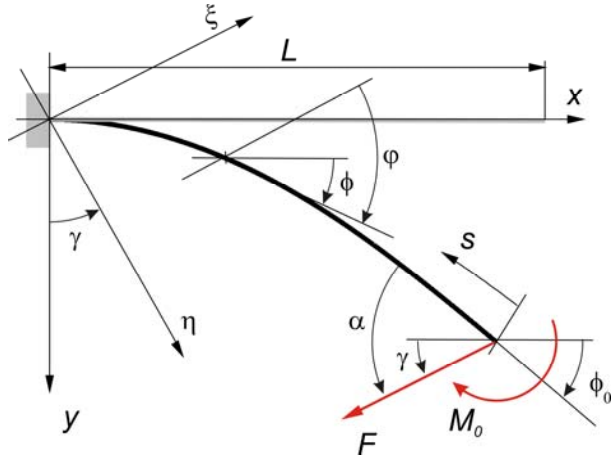

**Figure 1.** Geometry and load of the problem.

The equilibrium equations of the cantilever are ([54],pp 96)

$$H = -F\cos\gamma \qquad V = F\sin\gamma \tag{3}$$

$$\frac{dM}{ds} = -H\sin\phi + V\cos\phi = F\sin(\phi + \gamma) \tag{4}$$

where $H$ and $V$ are the horizontal and the vertical component of internal force, $M(s)$ is the bending toeque, $F \geq 0$ is magnitude of terminal force and $\gamma$ is the angle between the x-axis and the direction of force. If $\alpha$ is the angle between tangent to the cantilever base curve at the free point and the direction of the terminal force, and $\phi_0$ is the free end tangent angle then

$$\gamma = \alpha - \phi_0 \qquad \phi_0 \equiv \phi(0) \tag{5}$$

We assume that the torque and the curvature are related by the Bernoulli-Euler constitutive equation

$$M = EI\kappa \tag{6}$$

where $EI$ is assumed to be a positive constant which represents flexural rigidity of the cantilever. This equation together with the system of differential equations (1), (2) and (4) constitutes a complete set of equations for unknowns $x(s)$, $y(s)$, $\phi(s)$, $\kappa(s)$ and $M(s)$. The task is to solve these equations subject to the following boundary conditions

$$x(L) = y(L) = 0 \quad \phi(L) = 0 \quad \text{(clamped end)} \tag{7}$$

$$M(0) = M_0 \quad \text{(free end)} \tag{8}$$

where $M_0$ is applied torque.

*Non-dimensional form of equations.* The equation of the problem contains five parameters: $F$, $M_0$, $EI$, $L$ and $\gamma$. We reduce this number by introducing the load parameter $\omega$ defined by

$$\omega^2 \equiv \frac{FL^2}{EI} \tag{9}$$

the dimensionless curvature of cantilever free end $\kappa_0$ defined by

$$\kappa_0 \equiv \frac{M_0 L}{EI} \tag{10}$$

and the following normalization of geometric variables

$$\frac{s}{L} \to s \in [0,1] \quad \frac{x(s)}{L} \to x(s) \quad \frac{y(s)}{L} \to y(s) \quad L\kappa(s) \to \kappa(s) \tag{11}$$

We obtain some future simplification of the equations by introducing a new local coordinate system $O\xi\eta$ that is with respect to coordinate system $Oxy$ rotated by the angle $-\gamma$ so the line of action of applied force becomes parallel to $\xi$ axis (see Figure 1). In the new coordinate system the coordinates $\xi(s)$ and $\eta(s)$ of the cantilever base curve are

$$\xi(s) = x(s)\cos\gamma - y(s)\sin\gamma \quad \eta(s) = x(s)\sin\gamma + y(s)\cos\gamma \tag{12}$$

and the angle $\varphi(s)$ between tangent to the cantilever base curve and the $\xi$-axis is

$$\varphi(s) = \phi(s) + \gamma \tag{13}$$

The differential equations that describe cantilever shape in $O\xi\eta$ are obtained by differentiating (12) with respect to $s$, and then using (1) and (13). This gives

$$\frac{d\xi}{ds} = -\cos\varphi \quad \frac{d\eta}{ds} = -\sin\varphi \tag{14}$$

and from (7)$_{1,2}$ and (12) the associated boundary conditions are

$$\xi(1) = 0 \quad \eta(1) = 0 \tag{15}$$

Further, by using (6), (9) and (13) the differential equations (2) and (4) become

$$\frac{d\varphi}{ds}=-\kappa \qquad \frac{d\kappa}{ds}=\omega^2 \sin\varphi \qquad (16)$$

and from (7)$_3$ and (8) the associated boundary conditions are

$$\varphi(0)=\alpha \qquad \kappa(0)=\kappa_0 \qquad (17)$$

By selected sense of integration and rotation of the coordinate system we thus transform the original two point boundary value problem into a three parameter initial value problem (16)-(17) for unknowns $\varphi=\hat{\varphi}(s;\alpha,\omega,\kappa_0)$ and $\kappa=\hat{\kappa}(s;\alpha,\omega,\kappa_0)$. Once these functions have been determined, we can obtain the coordinates of a deformed cantilever base curve $\xi=\hat{\xi}(s;\alpha,\omega,\kappa_0)$ and $\eta=\hat{\eta}(s;\alpha,\omega,\kappa_0)$ by integration of (14) subject to boundary conditions (15). Further, from (13) the tangent angle $\phi$ is

$$\phi(s)=\varphi(s)-\gamma \qquad (18)$$

and by solving (12) for $x(s)$ and $y(s)$, we finally get

$$x(s)=\xi(s)\cos\gamma+\eta(s)\sin\gamma \qquad y(s)=-\xi(s)\sin\gamma+\eta(s)\cos\gamma \qquad (19)$$

Because the right hand sides of equations (16) and (14) are continuous functions, the existence theorem for ordinary differential equations guarantees the uniqueness of the solution for given initial conditions ([55],pp 144).

We see that the shape of a deformed cantilever depends on parameters $\alpha$, $\omega$ and $\kappa_0$ while its spatial position depends on $\gamma$. The relations between $\alpha$, $\omega$, $\kappa_0$ and $\gamma$ is obtain from (18). When $s=1$ we must have $\phi(1)=0$ and therefore by (18)

$$\gamma=\hat{\phi}(1;\alpha,\omega,\kappa_0)=\hat{\gamma}(\alpha,\omega,\kappa_0) \qquad (20)$$

This relation is fundamental and it allows us to define various types of problems. Some of them will be discussed in section 7. Until then we will assume that the given parameters are $\alpha$, $\omega$, $\kappa_0$.

*Symmetry*. If we in initial conditions (17) replace $\alpha$ by $-\alpha$ and $\kappa_0$ by $-\kappa_0$ then equations (14), (16), (18) and (19) imply the following symmetry

$$\kappa(s;-\alpha,-\kappa_0)=-\kappa(s;\alpha,\kappa_0) \qquad \varphi(s;-\alpha,-\kappa_0)=-\varphi(s;\alpha,\kappa_0)$$

$$\xi(s;-\alpha,-\kappa_0)=\xi(s;\alpha,\kappa_0) \qquad \eta(s;-\alpha,-\kappa_0)=-\eta(s;\alpha,\kappa_0) \qquad (21)$$

$$\phi(s;-\alpha,-\kappa_0)=-\phi(s;\alpha,\kappa_0) \qquad x(s;-\alpha,-\kappa_0)=x(s;\alpha,\kappa_0) \qquad y(s;-\alpha,-\kappa_0)=-y(s;\alpha,\kappa_0)$$

This shows that functions $\kappa(s)$, $\varphi(s)$, $\phi(s)$, $\eta(s)$ and $y(s)$ are odd and $\xi(s)$ and $x(s)$ are even functions of $\alpha$ and $\kappa_0$. The deformed shape of cantilever is thus symmetric with respect to $\xi$-axis and x-axis when $\alpha$ and $\kappa_0$ changes sign.

**3 Two special solutions**

*Trivial solution.* The initial value problem (16)-(17) has the following trivial solution ([54], pp 217). When $\kappa_0 = 0$ the boundary conditions (17) are satisfied by

$$\varphi(s) = \alpha \quad \text{and} \quad \kappa(s) = 0 \tag{22}$$

From (20) we then get $\gamma = \alpha$ while system (16) is reduced to $\sin\alpha = 0$, so we must have $\alpha = \pm n\pi$ where *n* is any integer. This gives two physical possibilities:

1. $\alpha = \gamma = 0$ (pure compression) and
2. $\alpha = \gamma = \pm\pi$ (pure extension).

In either case we from (13) obtain $\phi(s) = 0$. Therefore, the equations for cantilever shape (1) are reduce to $\frac{dx}{ds} = -1$ and $\frac{dy}{ds} = 0$ and these after integration under the boundary conditions (15) yield

$$x(s) = 1-s \quad \text{and} \quad y(s) = 0 \tag{23}$$

In words: for trivial solution the cantilever under arbitrary force remains straight. Future, because the solution of an initial value problem is unique, we conclude that in the cases when $\alpha = 0$ or when $\alpha = \pm\pi$ the only possible solution of the problem is trivial solution.

*Cantilever subject only to applied torque.* In this case $\omega = 0$ and $\gamma = 0$ so by (18) we have $\phi(s) = \varphi(s)$ while equations (16) are reduced to

$$\frac{d\varphi}{ds} = -\kappa \qquad \frac{d\kappa}{ds} = 0 \tag{23}$$

The integration of these equations under boundary conditions (17) and condition $\varphi(1) = 0$ yields

$$\varphi = \kappa_0(1-s) \qquad \kappa = \kappa_0 \tag{24}$$

where $\alpha = \kappa_0$. From (14), (15) and (19) we than have

$$x = \frac{\sin(\kappa_0(1-s))}{\kappa_0} \qquad y = \frac{1-\cos(\kappa_0(1-s))}{\kappa_0} \tag{25}$$

This is a well-known result which shows that a cantilever deforms into a circular arc laying on the circle

$$x^2 + \left(\frac{1}{\kappa_0} - y\right)^2 = 1/\kappa_0^2 \tag{26}$$

From (25) the coordinates of a cantilever free end are

$$x_0 \equiv x(0) = \frac{\sin \kappa_0}{\kappa_0} \qquad y_0 \equiv y(0) = \frac{1 - \cos \kappa_0}{\kappa_0} \tag{27}$$

In the special case when $x_0 = 0$ we from (27) have $\kappa_0 = n\pi$ $(n = \pm 1, \pm 2,...)$ and $y_0 \equiv y(0) = \frac{1-(-1)^n}{\pi n}$. The underlying circle is in this case is $x^2 + (1/n\pi - y)^2 = 1/(n\pi)^2$. The cantilever deforms to $n$ overlapping circles when $n$ is even; i.e., when $\kappa_0 = 2m\pi$ $(m = 1, 2,...)$. When $n \to \infty$ the cantilever reduces to a point.

From now on we will assume that $\omega > 0$.

**4 General solution**

The procedure of solution of initial value problem (16)-(17) is well known [56, 57] and therefore we shall here, for completeness, reproduce only the essential steps. In first step wet, by standard transformation $\frac{d\kappa}{ds} = \frac{d\kappa}{d\varphi}\frac{d\varphi}{ds} = \frac{d}{d\varphi}\left(\frac{\kappa^2}{2}\right)$ and integration under boundary conditions (17), obtain the first integral

$$\frac{d\varphi}{ds} = \sqrt{2\omega^2(\cos\varphi - \cos\alpha) + \kappa_0^2} = 2\omega\sqrt{\sin^2\frac{\alpha}{2} - \sin^2\frac{\varphi}{2} + \left(\frac{\kappa_0}{2\omega}\right)^2} \tag{28}$$

We will now discuss several cases of solution of this equation.

*Force dominant case.* This is the case when $\sin^2\frac{\alpha}{2} + \left(\frac{\kappa_0}{2\omega}\right)^2 < 1$. By introducing a new variable $\psi(s)$ defined by

$$\sin\frac{\varphi}{2} = k\sin\psi \tag{29}$$

where $k$ is the modulus defined by

$$k \equiv \sqrt{\sin^2\frac{\alpha}{2} + \left(\frac{\kappa_0}{2\omega}\right)^2} \tag{30}$$

we transform equation (28) into the following form

$$\frac{d\psi}{\omega ds} = \sqrt{1 - k^2 \sin^2\psi} \tag{31}$$

If we further set

$$u \equiv \sin\psi \qquad (32)$$

then equation (31) takes the Jacobi normal form

$$\frac{du}{\omega\, ds} = \sqrt{1-u^2}\sqrt{1-k^2 u^2} \qquad (33)$$

The general solution of this equation is

$$u(s) = \operatorname{sn}(\omega s + C, k) \qquad (34)$$

where $C$ is the constant of integration and $sn$ is the Jacobi elliptic sine function. From this we by using (32), (29) and (16)$_1$ find the solution of the problem

$$\varphi(s) = 2\sin^{-1}\left[k\operatorname{sn}(\omega s + C, k)\right] \qquad (35)$$

$$\kappa(s) = -2\omega k \operatorname{cn}(\omega s + C, k) \qquad (36)$$

where $cn$ is Jacobi's elliptic cosine function. When $s = 1$ equation (35) yields the explicit expression for (20)

$$\gamma = 2\sin^{-1}\left[k\operatorname{sn}(\omega + C, k)\right] \qquad (37)$$

The graph of this function for the special case when $\kappa_0 = 5$ is shown in Figure 2.

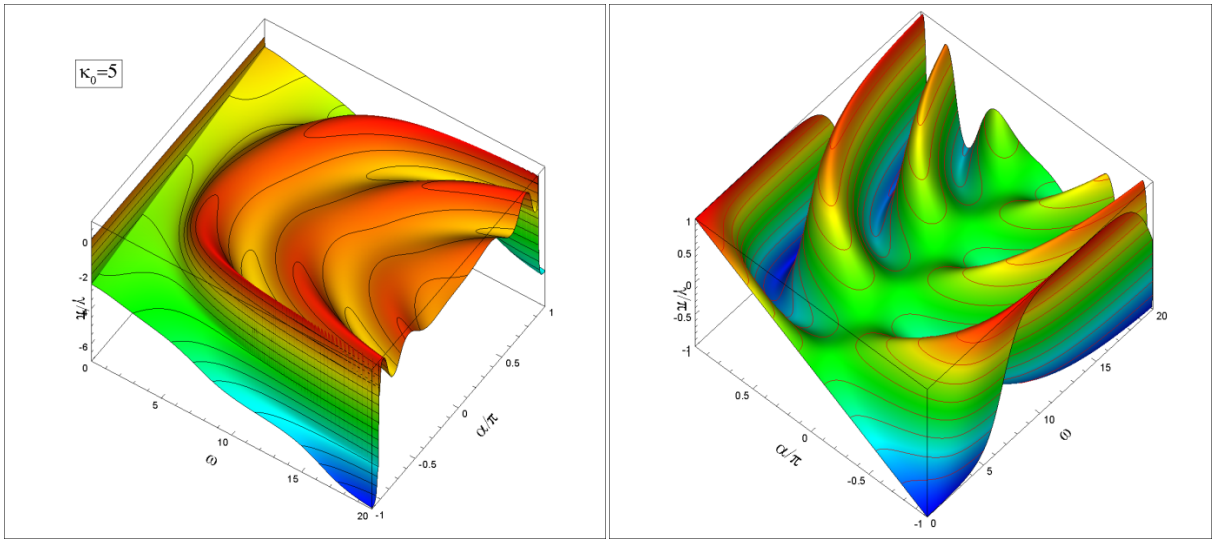

**Figure 2.** On left is the graph of $\gamma = \hat{\gamma}(\alpha, \omega, \kappa_0)$ given by Eq (37) when $\kappa_0 = 5$. On the right is the graph of $\gamma = \hat{\gamma}(\alpha, \omega)$ given by Eq (70). Note that at the end of interval $\gamma(\omega, \pm\pi) = \pm\pi$ but this is not displayed.

The constant of integration C is determined from initial conditions (17). By equating these conditions with values of (35) and (36) for $s=0$ we obtain two equations

$$\operatorname{sn}(C,k) = \frac{\sin(\alpha/2)}{k} \qquad -2\omega k \operatorname{cn}(C,k) = \kappa_0 \tag{38}$$

Inspection of the four possible combinations of signs of $\alpha$ and $\kappa_0$ yields the following expression for C

$$C = \begin{cases} \operatorname{sn}^{-1}\left(\dfrac{\sin(\alpha/2)}{k},k\right) & \kappa_0 < 0 \\[2ex] 2K - \operatorname{sn}^{-1}\left(\dfrac{\sin(\alpha/2)}{k},k\right) & \kappa_0 > 0 \end{cases} \tag{39}$$

where $K = K(k)$ is a complete elliptic integral of the first kind.

Now, in order to integrate equations (14) we by using (35) first express

$$\cos\varphi = 1 - 2\sin^2\frac{\varphi}{2} = 1 - 2k^2\operatorname{sn}^2(\omega s + C,k) = -1 + 2\operatorname{dn}^2(\omega s + C,k)$$
(40)
$$\sin\varphi = 2\sin\frac{\varphi}{2}\cos\frac{\varphi}{2} = 2\operatorname{sn}(\omega s + C,k)\sqrt{1 - k^2\operatorname{sn}^2(\omega s + C,k)} = 2k\operatorname{sn}(\omega s + C,k)\operatorname{dn}(\omega s + C,k)$$

where dn is the Jacobi elliptic delta function. The integral of these functions are ([57])

$$\int \cos\varphi\, ds = -s + \frac{2}{\omega}\int \operatorname{dn}^2(z+C,k)\, dz = -s + \frac{2}{\omega}\mathcal{E}(\omega s + C,k)$$
(41)
$$\int \sin\varphi\, ds = \frac{2}{\omega}\int \operatorname{sn}(z+C,k)\operatorname{dn}(z+C,k)\, dz = -\frac{2}{\omega}\operatorname{cn}(\omega s + C,k)$$

where $\mathcal{E}$ is Jacobi's epsilon function ([58],22.16.17,pp 562, [59], pp 517)

$$\mathcal{E}(z,k) \equiv \int_0^z \operatorname{dn}^2(t,k)\, dt \tag{42}$$

Instead of the Jacobi epsilon function we will use the Jacobi zeta function that is defined as ([58],22.16.32,pp 562)

$$Z(z,k) \equiv \mathcal{E}(z,k) - \frac{E}{K}z \tag{43}$$

where E is a complete Legandre elliptic integral of the second kind. Introducing the Jacobi zeta function has several advantages. First: it clearly separates the periodic part of the solution from its non-periodic part, second: the periodic part becomes bounded, and third: the Jacobi zeta function is part of the Maple program while the Jacobi epsilon function is not.

By substituting (41) into (14) and then applying the boundary conditions (15), we obtain the following parametric equations of a deformed cantilever base curve in local coordinate system

$$\xi(s) = \left(\frac{2E}{K} - 1\right)(1-s) + \frac{2}{\omega}\left[Z(\omega+C,k) - Z(\omega s+C,k)\right]$$

$$\eta(s) = -\frac{2k}{\omega}\left[\operatorname{cn}(\omega+C,k) - \operatorname{cn}(\omega s+C,k)\right]$$

(44)

Before proceeding we will derive some inequalities based on the fact that the trigonometric and Jacobian elliptic functions oscillate between -1 and 1. Assume that $\alpha \geq 0$. Then from (35) we obtain the interval for tangent angle

$$|\varphi(s)| \leq \alpha \tag{45}$$

We note that here we take into account that $\alpha$ is physically bounded to interval $[-\pi, \pi]$ and so we must in (35) take the principal value of function $\sin^{-1}$. In this way expression (37) gives the unique value of $\gamma$. In particular we for $s=0$ obtain (see also [48], Eq 11)

$$|\gamma| \leq \alpha \tag{46}$$

From this inequality and equation (5) follows the range for the free end tangent angle

$$0 \leq \phi_0 \leq 2\alpha \tag{47}$$

The range for a cantilever curvature follows from (36) and is

$$|\kappa(s)| \leq 2\omega \tag{48}$$

<u>Torque dominant case.</u> This is the case when $\sin^2\frac{\alpha}{2} + \left(\frac{\kappa_0}{2\omega}\right)^2 \geq 1$. By setting $\psi = \varphi/2$ we from (28) obtain the equation

$$\frac{d\psi}{\omega ds} = k\sqrt{1 - k^{-2}\sin^2\psi} \tag{49}$$

which, after performing similar transformations as in the previous case, leads to the solution of (16) in the form

$$\varphi(s) = -2\operatorname{sgn}(\kappa_0)\operatorname{am}(k\omega s + C, k^{-1}) \tag{50}$$

$$\kappa(s) = 2\operatorname{sgn}(\kappa_0)\omega k \operatorname{dn}(k\omega s + C, k^{-1}) \tag{51}$$

where $\operatorname{am}(x) = \arcsin(\operatorname{sn}(x))$ is Jacobi's amplitude function and constant $C$ is given by

$$C = -\operatorname{sgn}(\kappa_0)\operatorname{sn}^{-1}\left(\sin\frac{\alpha}{2}, k^{-1}\right) \tag{52}$$

The choice of signs in these equations deserve a little explanation. Since $dn$ is always positive, the sign of $\kappa$ is determined from the initial condition. Now assume that $\kappa_0 > 0$. By (16)$_1$ the derivation of $\varphi$ should be negative, therefore (50) must be a negative signet. But for $\alpha > 0$ we must have $\varphi(0) = -2\operatorname{am}(C, k^{-1}) > 0$ and therefore $C$ should be negative. The reasoning for $\kappa_0 < 0$ is similar. The explicit form of (20) is in this case from (50)

$$\gamma = -2\operatorname{sgn}(\kappa_0)\operatorname{am}(k\omega + C, k^{-1}) \tag{53}$$

In order to integrate equations (14) we first note that $\cos\operatorname{am}(x) = \operatorname{cn} x$ and $\sin\operatorname{am}(x) = \operatorname{sn} x$ hence we by using (50) obtain

$$\cos\varphi = 1 - 2\operatorname{sn}^2(k\omega s + C, k^{-1}) = 1 - 2k^2 - 2k^2\operatorname{dn}^2(k\omega s + C, k^{-1})$$

$$\sin\varphi = -2\operatorname{sgn}(\kappa_0)\operatorname{sn}(k\omega s + C, k^{-1})\operatorname{cn}(k\omega s + C, k^{-1}) \tag{54}$$

and therefore

$$\int\cos\varphi\,ds = s - \frac{2}{k\omega}\int\operatorname{sn}^2(z + C, k^{-1})dz = (1 - 2k^2)s + \frac{2k}{\omega}\mathcal{E}(k\omega s + C, k^{-1})$$

$$\int\sin\varphi\,ds = -\frac{2\operatorname{sgn}(\kappa_0)}{\omega k}\int\operatorname{sn}(z + C, k^{-1})\operatorname{cn}(z + C, k^{-1})dz = \operatorname{sgn}(\kappa_0)\frac{2k}{\omega}\operatorname{dn}(k\omega s + C, k^{-1}) \tag{55}$$

By using the above integrals under conditions (15) we from (14) obtain the parametric equation of a cantilever base curve in the form

$$\xi(s) = \left[2k^2\left(\frac{E(k^{-1})}{K(k^{-1})} - 1\right) + 1\right](1 - s) + \frac{2k}{\omega}\left[Z(k\omega + C, k^{-1}) - Z(k\omega s + C, k^{-1})\right]$$

$$\eta(s) = \operatorname{sgn}(\kappa_0)\frac{2k}{\omega}\left[\operatorname{dn}(k\omega + C, k^{-1}) - \operatorname{dn}(k\omega s + C, k^{-1})\right] \tag{56}$$

Consider now the special case when $\kappa_0/\omega \to \infty$ and therefore $k^{-1} \to 0$. By using the Maclaurin series of functions *dn* ([58], 22.10.6, pp 559) and definition (43) we obtain the following expansion of Z with respect to *k*

$$Z(z,k) = -\left(1 - \frac{E(k)}{K(k)}\right)z - \frac{k^2}{4}(2z - \sin 2z) + O(k^4) \tag{57}$$

Further we have

$$\lim_{k \to \infty} 2k^2 \left(\frac{E(1/k)}{K(1/k)} - 1\right) = -1 \tag{58}$$

From (50),(51) and (56) we can now deduce

$$\varphi(s) \to -\kappa_0 s + \alpha \qquad \kappa(s) \to \kappa_0 \tag{59}$$

$$\xi(s) \to \frac{\sin(\kappa_0 - \alpha) - \sin(\kappa_0 s - \alpha)}{\kappa_0}$$

$$\eta(s) \to \frac{\cos(\kappa_0 - \alpha) - \cos(\kappa_0 s - \alpha)}{\kappa_0} \tag{60}$$

and from (18) and (19)

$$\phi(s) \to \kappa_0(1-s) \tag{60}$$

$$x(s) \to \frac{\sin[\kappa_0(1-s)]}{\kappa_0} \qquad y(s) \to \frac{1 - \cos[\kappa_0(1-s)]}{\kappa_0} \tag{61}$$

These show that the solution approaches the solution of a cantilever subject only to a torque.

<u>Case when *k* =1</u>. The condition in case $\alpha = \pi$ implies $\kappa_0 = 0$ and this is covered by a trivial solution. For $|\alpha| < \pi$ we from (30) have

$$\kappa_0 = 2\omega \cos\frac{\alpha}{2} \tag{62}$$

where we assume that $\kappa_0 > 0$. From (50), (51), (53) and (56) we then obtain

$$\varphi(s) = -2\sin^{-1}[\tanh(\omega s + C)] \qquad \kappa(s) = \frac{2\omega}{\cosh(\omega s + C)} \tag{63}$$

$$\xi(s) = -1 + s + \frac{2}{\omega}\left[\tanh(\omega + C) - \tanh(\omega s + C)\right]$$

$$\eta(s) = \frac{2}{\omega}\left[\frac{1}{\cosh(\omega + C)} - \frac{1}{\cosh(\omega s + C)}\right]$$

(64)

where, from (52)

$$C = -\tanh^{-1}\left(\sin\frac{\alpha}{2}\right) \tag{65}$$

*Case when* $\kappa_0 = 0$. In this special case we from (30) have

$$k \equiv \sin\frac{\alpha}{2} \tag{66}$$

and therefore (39) becomes

$$C = \operatorname{sn}^{-1}(1) = K(k) \tag{67}$$

From (35) and (36) we have

$$\varphi(s) = 2\sin^{-1}\left[k\operatorname{sn}(\omega s + K, k)\right] = 2\sin^{-1}\left[k\frac{\operatorname{cn}(\omega s, k)}{\operatorname{dn}(\omega s, k)}\right] \tag{68}$$

$$\kappa(s) = -2\omega k\operatorname{cn}(\omega s + K, k) = 2\omega k k'\frac{\operatorname{sn}(\omega s, k)}{\operatorname{dn}(\omega s, k)} \tag{69}$$

where $k' = \sqrt{1-k^2}$. In particular, we from (68) for $s = 1$ obtain

$$\gamma = 2\sin^{-1}\left[k\operatorname{sn}(\omega + K, k)\right] = 2\sin^{-1}\left[k\frac{\operatorname{cn}(\omega, k)}{\operatorname{dn}(\omega, k)}\right] \tag{70}$$

The graph of this function is shown in Figure 2. The parametric equations for a deformed cantilever shape in a local coordinate system follows from (44) and are

$$\xi(s) = \left(\frac{2E}{K} - 1\right)(1-s) + \frac{2}{\omega}\left[Z(\omega+K,k) - Z(\omega s+K,k)\right]$$

$$= \left(\frac{2E}{K} - 1\right)(1-s) + \frac{2}{\omega}\left\{Z(\omega,k) - Z(\omega s,k) - k^2\left[\frac{sn(\omega,k)cn(\omega k)}{dn(\omega,k)} - \frac{sn(\omega s,k)cn(\omega s,k)}{dn(\omega s,k)}\right]\right\} \quad (71)$$

$$\eta(s) = -\frac{2k}{\omega}\left[cn(\omega+K,k) - cn(\omega s+K,k)\right] = \frac{2kk'}{\omega}\left[\frac{sn(\omega,k)}{dn(\omega,k)} - \frac{sn(\omega s,k)}{dn(\omega s,k)}\right]$$

The expanded form of the solution is useful for calculation purposes since it avoids numerical problems for the special case when $\alpha = \pi$ and therefore $k=1$ and $K(1) = \infty$. For this case since $sn(z,1) = Z(z,1) = \tanh z$ and $cn(z,1) = dn(z,1) = 1/\cosh z$ ([57], pp 16) we from (68)-(71) obtain a trivial solution. When $\alpha = 0$ then $k=0$ and therefore $sn(z,0) = \sin z$, $cn(z,0) = \cos z$, $dn(z,0) = 1$, $Z(z,0) = 0$ and $E = K = \pi/2$ ([57], pp 15). Substituting these values into (68)-(71) also yields a trivial solution.

The present solution given by Eq (68), (69),(70) and (71) contains as an argument a complete elliptic integral $K$ while the solution functions given in [48] contain as an argument the incomplete elliptic integral of the first kind. Also in the present solution the direction of force $\gamma$ is given explicitly by (70) while in [48] (Eq 9) the explicit expression for load parameter $\omega$ is given. As we will show this expression is the solution of equation (68) when $\omega$ is taken as the unknown.

## 5 Numerical examples

For purpose of numerical calculations we wrote a computer program where for calculation of Jacobian elliptic functions we use a slightly modified subroutine JELP from [60] and for calculation of Legandre elliptic integrals and Z-function routines from ACM Algorithm 577 ([25]). All numerical computations were executed in a double precision numerical model.

In Table 1 are some comparisons of results of calculations obtained by our program and calculations obtained by the Maple program, where the number of digits was set to 14. As can be seen the results of calculations match to 11 digits.

Tables 2,3 and 4 present comparisons of results obtained by the present method and the numerical solution of the problem. By following the idea of Shvartzman ([41]) we executed the integration in two steps. First the initial value problem (16)-(17) is solved. The result of integration are a cantilever fixed end point curvature $\kappa_1 = \kappa(1)$ and force angle $\gamma = \varphi(1)$. With these data and by changing the orientation with $s \rightarrow 1-s$ the equations (1), (2), (4), (6) and (9) become

$$\frac{dx}{ds} = \cos\phi \quad \frac{dy}{ds} = \sin\phi \quad \frac{d\phi}{ds} = \kappa \quad \frac{d\kappa}{ds} = -\omega^2 \sin(\phi+\gamma) \quad s \in [0,1] \quad (72)$$

and the associated initial conditions are

$$x(0) = y(0) = 0 \quad \phi(0) = 0 \quad \kappa(0) = \kappa_1 \tag{73}$$

This is also an initial value problem that can be solved numerically without iteration. We note that our second step is different from one purposed in [41] where the Simpson integration is used to obtain a beam shape and this presumably required storing of the data for $\phi(s)$ from the first integration step. For the numerical integration we use subroutine *dopri5* which implements an explicit Runge-Kutta method of order 4-5 with stepsize control [61] . Results of calculation of beam shape show that when absolute and relative error of calculation was set to $10^{-7}$ the results of analytical and numerical integration agree to 6 digits. Results of calculations in Table 3 and 4 were conducted by setting absolute and relative error of calculation to $10^{-9}$. For these cases, results match to within 8 digits. The shapes of a deformed cantilever shown in Figure 3 and Figure 4 correspond to cases presented in Table 3 and 4, respectively.

In Table 5 we present the comparison of results of calculation of tip angle and tip coordinates for $\alpha = 90^0$ obtained by several authors. The results obtained by Shvartsman [41] and Nageswara and Venkateswara [44] are identical with those obtained by the present calculation while the discrepancy with results given by Mutyalarao et al. [43] is at most by 4%. This discrepancy can be explained by the fact that the authors for integration used fourth order Runge-Kutta method with fixed integration step 0.001.

**Table 1.** Comparison of results of calculated free end coordinates $(x_0, y_0)$, free end tangent angle $\phi_0$ and root curvature $\kappa_1$ for $\alpha = 90^0$, $\kappa_0 = 0$ and various values of load parameter $\omega$

| $\omega$ | $x_0$ | $y_0$ | $\phi_0/\pi$ | $\kappa_1/\omega$ | Difference with Maple $\times 10^{-12}$ | | | |
|---|---|---|---|---|---|---|---|---|
| | | | | | $\Delta x_0$ | $\Delta y_0$ | $\Delta \phi_0$ | $\Delta \kappa_1$ |
| 1 | 0.935645669481 | 0.320641994675 | 0.157844984090 | 0.975510043970 | 0.33 | -0.12 | -0.14 | 0.44 |
| 5 | 0.461585556493 | 0.102962763465 | 0.740348529768 | -1.206829444100 | -0.18 | -0.42 | -0.12 | -0.10 |
| 10 | 0.195867673290 | 0.457111103310 | 0.801395679728 | 1.080955458240 | -0.33 | -0.05 | -0.54 | 3.40 |
| 50 | 0.457047600338 | 0.010527023263 | 0.527000611134 | -1.411668886800 | -0.22 | -0.30 | 0.24 | 2.70 |
| 100 | 0.002095409686 | 0.457488959373 | 0.997706928372 | 0.120031951352 | -0.03 | 0.04 | -0.32 | -1.53 |

**Table 2.** Comparison of results of calculated coordinates $(x,y)$, tangent angle $\phi$ and curvature $\kappa$ when $\omega=10$, $\kappa_0=0$ and $\alpha=90^0$ and various values of *s*.

| s | x | y | $\phi/\pi$ | $\kappa/\omega$ | Difference with DOPRI5 $\times 10^{-6}$ | | | |
|---|---|---|---|---|---|---|---|---|
| | | | | | $\Delta x_0$ | $\Delta y_0$ | $\Delta \phi_0$ | $\Delta \kappa_0$ |
| 0.0 | 0.1958677 | 0.4571111 | 0.8013957 | 0.0000000 | -0.04 | -0.47 | 0.23 | 0.00 |
| 0.2 | 0.2514655 | 0.2948450 | 0.2359386 | 1.3992359 | -0.16 | -0.42 | -0.54 | 0.07 |
| 0.4 | 0.0729499 | 0.3363053 | -0.1850488 | -0.2917977 | -0.12 | -0.14 | -0.34 | 0.13 |
| 0.6 | -0.0809303 | 0.2925161 | 0.4923433 | -1.2848374 | -0.05 | -0.07 | 0.01 | -0.37 |
| 0.8 | 0.0431836 | 0.1453901 | 0.7472228 | 0.5820115 | -0.06 | 0.00 | 0.11 | -0.36 |
| 1.0 | 0.0000000 | 0.0000000 | 0.0000000 | 1.0809555 | 0.00 | 0.00 | 0.00 | 0.72 |

**Table 3.** Comparison of results of calculated free end coordinates $(x_0, y_0)$, free end tangent angle $\phi_0$ and root curvature $\kappa_1$ for $\alpha=90^0$ $\omega=5$ and various values of free end curvature $\kappa_0$.

| $\kappa_0$ | $x_0$ | $y_0$ | $\phi_0/\pi$ | $\kappa_1/\omega$ | Difference with *dopri5* $\times 10^{-9}$ | | | |
|---|---|---|---|---|---|---|---|---|
| | | | | | $\Delta x_0$ | $\Delta y_0$ | $\Delta \phi_0$ | $\Delta \kappa_1$ |
| 5 | 0.446415423 | -0.313062861 | 0.710386283 | -8.029354398 | 0.18 | 0.32 | 1.27 | 2.60 |
| 7.5 | 0.113376313 | 0.315858110 | 1.802376703 | 5.211869664 | -1.81 | 0.93 | 0.45 | 28.96 |
| 10 | 0.015288729 | 0.084345999 | 3.183130184 | 8.531987832 | -0.51 | 0.01 | 0.41 | 3.08 |
| 15 | 0.006704371 | 0.073344605 | 4.792504287 | 15.979160875 | 0.18 | 0.41 | 0.07 | 3.35 |
| -5 | -0.272119971 | -0.055314289 | 1.145431732 | 1.715272828 | 1.57 | -1.71 | 0.32 | 3.83 |
| -7.5 | -0.352524470 | 0.192348943 | -1.728937144 | -9.688433940 | 2.83 | -0.52 | 1.20 | -9.31 |
| -10 | 0.110036846 | -0.322462967 | -2.850046317 | -8.792437532 | -0.04 | 0.33 | 0.05 | 0.96 |
| -15 | 0.105702957 | -0.134393526 | -4.673781913 | -13.500678911 | -0.81 | -0.07 | 0.11 | -3.11 |

**Table 4.** Comparison of results of calculated free end coordinates $(x_0, y_0)$, free end tangent angle $\phi_0$ and root curvature $\kappa_1$ for case $k=1$ and various values of $\alpha$ and $\omega$. For analytical calculation Eqs (63) and (64) were used.

| $\alpha$ deg | $\omega$ | $x_0$ | $y_0$ | $\phi_0/\pi$ | $\kappa_1/\omega$ | Difference with *dopri5* $\times 10^{-9}$ | | | |
|---|---|---|---|---|---|---|---|---|---|
| | | | | | | $\Delta x_0$ | $\Delta y_0$ | $\Delta \phi_0$ | $\Delta \kappa_1$ |
| 90 | 1 | 0.498143105 | 0.708358790 | 0.575343418 | 1.986009816 | -0.19 | -0.16 | 0.12 | 0.35 |
| | 5 | 0.334240815 | 0.248627803 | 1.479290239 | 0.325250796 | -1.02 | 1.81 | 1.68 | 38.28 |
| -90 | 1 | 0.880845859 | 0.386135102 | 0.307464029 | 0.595690757 | -0.07 | -0.27 | 0.01 | -0.16 |
| | 5 | 0.885926529 | 0.270736769 | 0.496446463 | 0.055818546 | 2.55 | -3.75 | -0.81 | -13.24 |

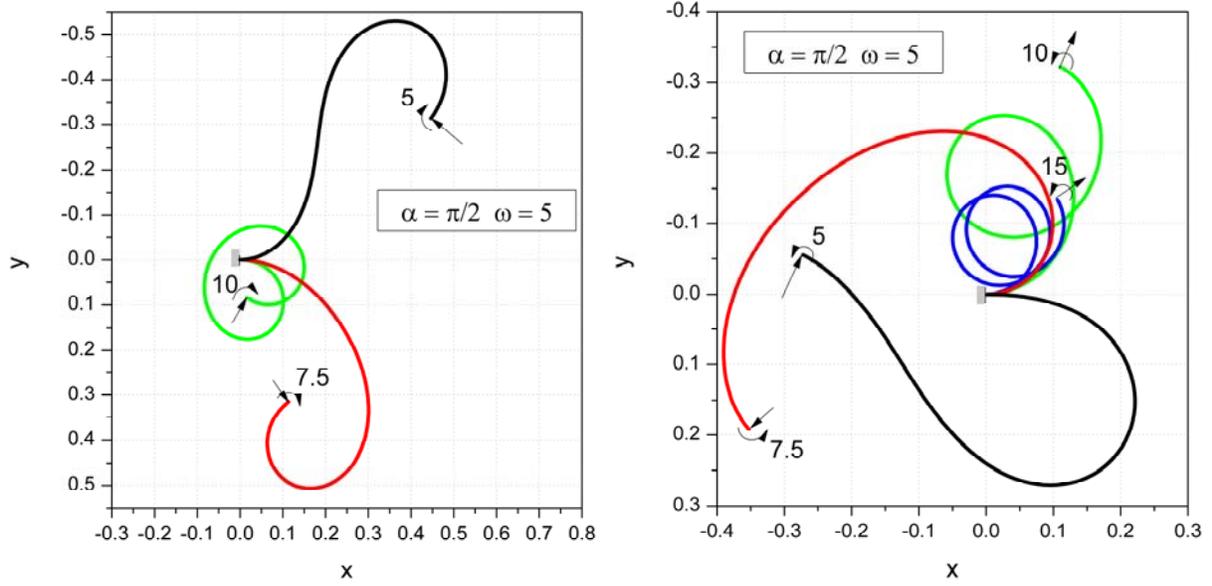

**Figure 3.** Cantilever shapes for various values of free end curvature $\kappa_0$ when $\alpha = 90^0$ and $\omega = 5$.

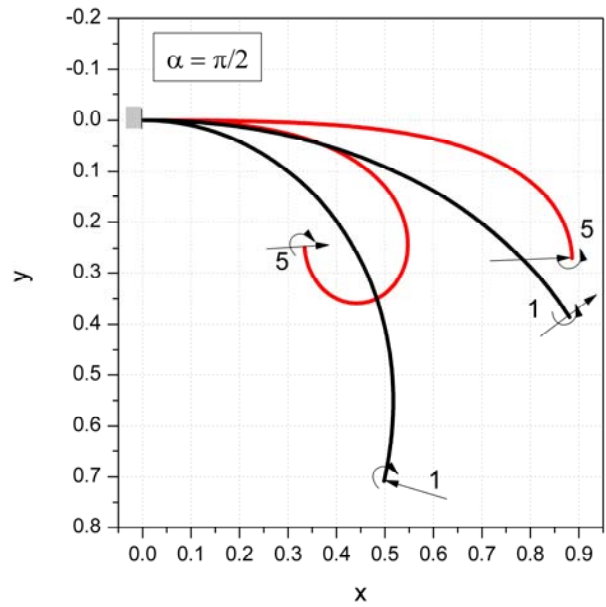

**Figure 4.** Cantilever shapes when $\alpha = 90^0$, $\kappa_0 = \sqrt{2}\omega$ and various values of $\omega$.

**Table 5.** Comparison of results of calculated of free end tangent angle $\phi_0$ and free end coordinates $(x_0, y_0)$ for $\alpha = 90^0$ obtained by other authors. Superscript * indicates input data that the authors used for calculation.

| $\omega^2$ | $\phi_0$ (deg) | $x_0$ | $y_0$ | Present | | | Reference |
|---|---|---|---|---|---|---|---|
| | | | | $\phi_0$ (deg) | $x_0$ | $y_0$ | |
| 0.7010 | 20* | 0.9678 | 0.2292 | 20.000435 | 0.967807 | 0.229225 | |
| 2.1755 | 60* | 0.7312 | 0.6082 | 60.001213 | 0.731196 | 0.608173 | [44] |
| 4.9872 | 120* | 0.1742 | 0.7810 | 119.999711 | 0.174160 | 0.781038 | |
| 2* | 55.48 | 0.7674 | 0.5738 | 55.475997 | 0.767362 | 0.573839 | |
| 13.75* | 180 | 0.0000 | 0.4570 | 180.000000 | -0.000014 | 0.456953 | [41] |
| 36* | 55.64 | 0.2855 | -0.4546 | 55.629426 | 0.285341 | -0.454598 | |
| 220.006 | 0* | -0.000060 | -0.457523 | 0.000000 | 0.000000 | -0.456947 | [43] Table 2 |
| 190.565 | 30* | 0.175045 | -0.450477 | 30.000709 | 0.176352 | -0.450321 | |
| 79.054 | 60* | 0.456241 | -0.086106 | 59.999350 | 0.451377 | -0.089691 | |

**6 Analysis of deformed cantilever base curve**

In this section we will give a detailed analysis of a deformed cantilever base line curve where the ultimate goal is classification of its possible forms. Historically such analysis was first given by Euler ([9],pp 199-213) using only integrals and functions series expansion. He showed that a deformed cantilever is a part of an infinite periodic curve called elastic. Later the analysis was by using Jacobi's elliptic functions given by Love ([62],pp 386-387) and more recently by Antman ([54],pp 98-100), who provides only qualitative analysis based on the phase portrait of (16) in $(\varphi, \kappa)$ plane. All these authors consider only a rod subject to a force. More general consideration of possible shapes of elastica were given by Goss [10] and Schakov [63] where Goss provided also experimental verification of analytical results.

The analysis is based on determination of cantilever base curve inflection points. By definition the inflection point is a point where $\kappa(s) = 0$. The curvature has a relative extreme at points where $\frac{d\kappa}{ds} = 0$. Following Zakharov [48] we will call these points compression points (see Figure 5).

<u>Force dominant case. Inflectional elastic.</u> In what follows we will assume that $k \neq 0$ and the angle corresponded to *k* is

$$\alpha' = 2\sin^{-1} k \tag{74}$$

Clearly, when $\kappa_0 = 0$ then $\alpha' = \alpha$.

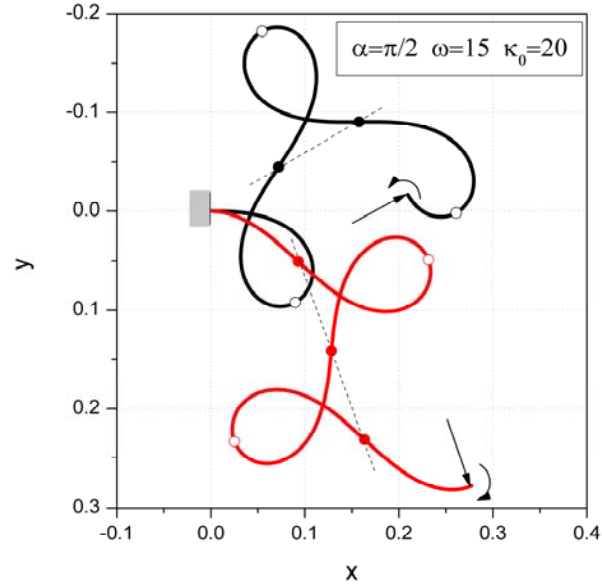

**Figure 5.** Inflection (solid circles) and compression points (hollow circles) for two cases with opposite sense of applied torque.

*Elastica.* In order to obtain the simplest form of parametric equations of elastica we introduce a new parameter $\sigma$ defined by

$$\sigma = \omega s + C \tag{75}$$

and translate the coordinate system into the point $s_0 = -C/\omega$. The parametric equations of elastica are then by (44)

$$\bar{\xi} \equiv \omega\left[\xi(s_0) - \xi(s_0 + \sigma/\omega)\right] = \left(\frac{2E}{K} - 1\right)\sigma + 2Z(\sigma, k)$$

$$\bar{\eta} \equiv \omega\left[\eta(s_0) - \eta(s_0 + \sigma/\omega)\right] = 2k\left[1 - \operatorname{cn}(\sigma, k)\right] \tag{76}$$

Since function *cn* has period $4K$ and function *Z* has period $2K$ the elastica is a periodic function with period $4K$ and its single wave is given by $0 \le \sigma < 4K$. We see that the shape of elastica depends only on *k* while $\omega$ plays a role of scale. In this new parameterization the tangent angle and the curvature are by (35) and (36)

$$\varphi(\sigma) = 2\sin^{-1}\left[k\operatorname{sn}(\sigma, k)\right] \qquad \bar{\kappa} \equiv \frac{\kappa(\sigma)}{\omega} = -2k\operatorname{cn}(\sigma, k) \tag{77}$$

and the inflection $\sigma_n$ and the compression points $\sigma'_n$ are at

$$\sigma_n = (2n+1)K \qquad \sigma'_n = 2nK \qquad n \in Z \tag{78}$$

By this we see that the arc length between successive inflection/compression points is 2*K* and the arc length of a single wave is 4*K*. Each elastic wave contains three compression points and two inflection points where the inflection point lies in the middle between successive compression points. All the inflection points lie on the line $\bar{\eta} = 2k$ and compression points alternate between the lines $\bar{\eta} = 0$ and $\bar{\eta} = 4k$. We note that all the lines of the form $\bar{\eta} = c$ are in space coordinates given by

$$x \sin\gamma + y \cos\gamma = \eta(s_0) - \frac{c}{\omega} = \frac{2k[1 - \text{cn}(\omega + C, k)] - c}{\omega} \tag{79}$$

At inflection points the value of the tangent angle is

$$\varphi(\sigma_n) = (-1)^{n-1} \alpha' \tag{80}$$

and at compression points the curvature is maximal

$$\bar{\kappa}_{max} = |\bar{\kappa}(\sigma'_n)| = 2k \tag{81}$$

*Dimensions.* The horizontal distance $\Delta\bar{\xi}_c$ and the vertical distance $\Delta\bar{\eta}_c$ (twice the amplitude of the wave) between two successive compression points are by (76) and (78)

$$\Delta\bar{\xi}_c \equiv \bar{\xi}(2K) - \bar{\xi}(0) = 2(2E - K) \qquad \Delta\bar{\eta}_c \equiv \bar{\eta}(2K) - \bar{\eta}(0) = 4k \tag{82}$$

As is known, when *k* increases the elastica begins to form loops. The loop extreme points in the horizontal direction are obtained from the condition $\frac{d\bar{\xi}}{d\sigma} = 1 - 2\text{dn}^2(\sigma, k) = 0$ providing that $k \geq \sqrt{2}/2$ or $\alpha' \geq \pi/2$. At the interval $0 \leq \sigma \leq 4K$ this leads to four values of parameter: $\zeta$, $2K - \zeta$, $2K + \zeta$, $4K - \zeta$ where

$$\zeta = \text{dn}^{-1}\left(\frac{\sqrt{2}}{2}, k\right) \tag{83}$$

This means that on each wave we have two possible loops. The loop width $\Delta\bar{\xi}_t$ and the vertical distance between successive extreme points $\Delta\bar{\eta}_t$ are

$$\Delta\bar{\xi}_t \equiv \bar{\xi}(\zeta) - \bar{\xi}(-\zeta) = 2\left(\frac{2E}{K} - 1\right)\zeta + 4Z(\zeta, k)$$

$$\Delta\bar{\eta}_t \equiv \bar{\eta}(2K - \zeta) - \bar{\eta}(\zeta) = 4k\,\text{cn}(\zeta, k) \tag{84}$$

Two examples of calculation by these formulas are shown in Figure 7 (cases a and b).

*Intersection points.* In order that elastica have self-intersection points we must have $\xi(\sigma_1)=\xi(\sigma_2)$, $\eta(\sigma_1)=\eta(\sigma_2)$ and $\sigma_1 \neq \sigma_2$. By (76) this requirement leads to system of two nonlinear algebraic equations for unknowns $\sigma_1$ and $\sigma_2$

$$0=\left(\frac{2E}{K}-1\right)(\sigma_2-\sigma_1)+2\left[Z(\sigma_2,k)-Z(\sigma_1,k)\right] \quad \text{cn}(\sigma_1,k)=\text{cn}(\sigma_2,k) \tag{85}$$

By introducing the new variables

$$\sigma_1=\zeta \text{ and } \sigma_2=-\zeta+4qK \quad \zeta\in(0,K) \quad (q=0,\pm 1,\ldots) \tag{86}$$

the second equation reduces to identity while the first becomes

$$\Phi_q(\zeta,k)\equiv\left(\frac{2E}{K}-1\right)(\zeta-2qK)+2Z(\zeta,k)=0 \quad (q=0,\pm 1,\ldots) \tag{87}$$

The graph of this relation is shown in Figure 6 (left). It is clear from this graph that for given $\alpha'$ we may obtain from zero up to an infinite number of intersection points. However only three values of $\alpha'$ are of some interest:

- value of $\alpha'$ when all the waves overlap each other ( $q=\infty$ )
- value of $\alpha'$ when the wave touches the adjacent ( $q=1$ ) and
- value of $\alpha'$ when the wave detaches ( $q=-1$ ).

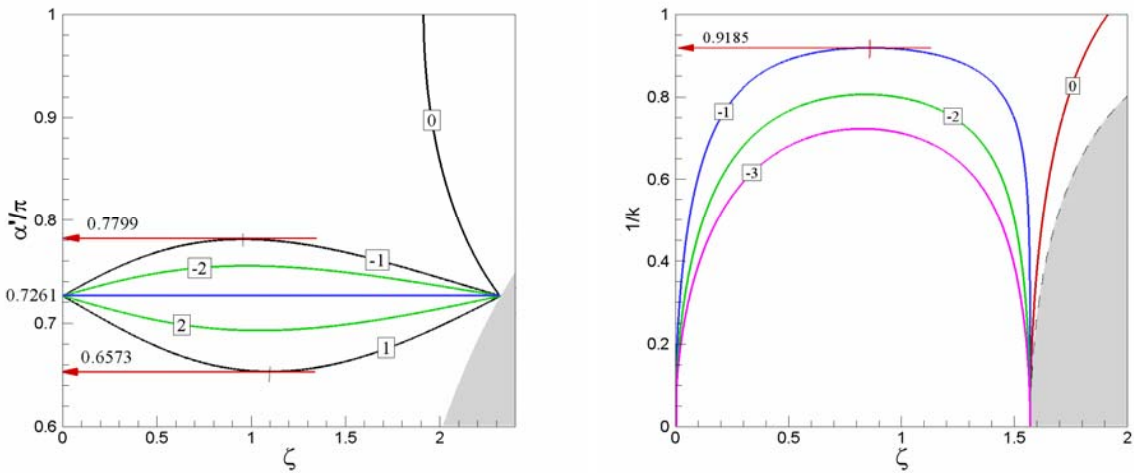

**Figure 6.** On the left is the graph of $\Phi_q\left(\zeta,\sin\dfrac{\alpha'}{2}\right)=0$ given by (87) for different values of $q$. On the right is the graph of relation (106). In dark regions $\zeta>K$

The solution for the overlapping case is obtained by setting $\zeta = K$ so equation (87) becomes the equation for unknown $k$ of the form $K(k) = 2E(k)$. The solution of this equation is $k \approx 0.9089086$ and this gives Euler's $\alpha' \approx 130^0\ 42'35.7''$ [1] (pp 154). This solution may also be obtained by the requirement that all the inflection and compression points coincide and therefore $(82)_1$ becomes $\Delta\overline{\xi}_c = 0$ which yields the equation $K = 2E$.

The value of $\alpha'$ for attachment and detachment are obtained by the solution of the system consisting of Eq (87) and the following equation

$$\frac{\partial \Phi_q}{\partial \zeta} = 1 - 2\text{dn}^2\zeta = 0 \qquad (88)$$

The solution of this equation is given by (83). Substituting (83) into (87) yields an equation for unknown $k$ which for $q = 1$ has the solution $k \approx 0.8550924$ and for $q = -1$ the solution $k \approx 0.9414031$. The first $k$ yields the attachment value $\alpha' \approx 117^0 32'23.6''$ and corresponding parameter $\zeta \approx 1.0997400$, and the second detachment value $\alpha' \approx 140^0 34'37.5''$ and parameter $\zeta \approx 0.9554893$. The shape and dimensions of elastica for these values of $\alpha'$ are shown in Figure 7 (cases a and b).

The parameter $\zeta$ which locates the wave self-intersection point for a given $k$ is the solution of the equation (87) when $q = 0$ that is the solution of equation

$$\left(\frac{2E}{K} - 1\right)\zeta + 2Z(\zeta, k) = 0 \qquad \zeta \in (0, K) \qquad (89)$$

At the given interval $Z \geq 0$ therefore we obtain the solution only when $2E < K$, that is, when $\alpha' > 130.7^0$. This can also be observed on the graph in Figure 6. Parameters that define intersection points on a single wave are then given by $\zeta$, $2K - \zeta$, $2K + \zeta$, $4K - \zeta$. The distances in coordinate directions between successive intersection points are given by (84) where $\zeta$ is the solution of (89).

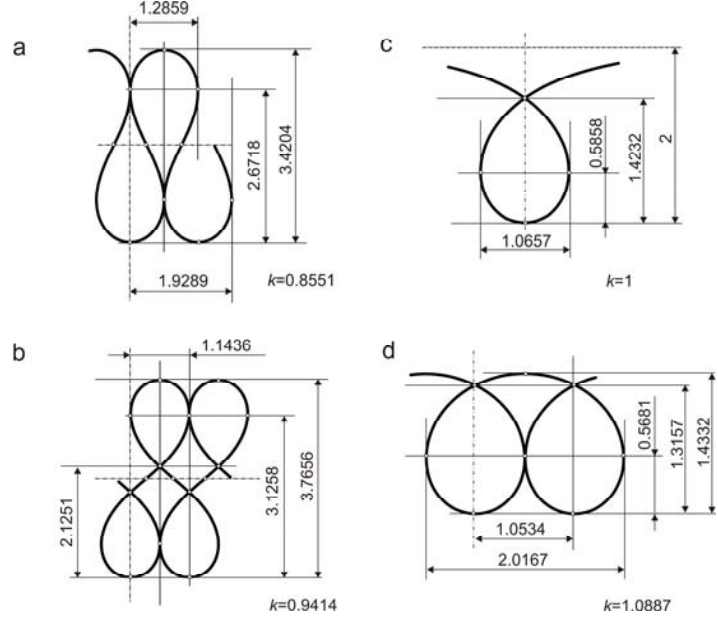

**Figure 7.** Some characteristic dimensions of elastica for various values of *k*. To obtain physical dimensions the values for cases *a*, *b* and *c* should be multiplied by $L/\omega$, while the values for case *d* should be multiplied by $L/k\omega$. The values for case *a* and *c* agree with those given by Goss [10] (Fig 4.16 and Fig 4.9)

*Cantilever.* In order to determine the number of inflection *m* and compression points *m'* on a cantilever we first note that the new parameterization of the end points of the cantilever are determined by

$$\sigma_0 = C \qquad \sigma_1 = C + \omega \tag{90}$$

Based on the definition of *C* given by (39) we distinguish three cases.

- *Case when $\kappa_0 < 0$*. In this case we have $0 < C \leq K$ so the number of inflection points *m* and the number of compression points *m'* are given by

$$m = 1 + \text{floor}\left(\frac{\omega - K + C}{2K}\right) \qquad m' = \text{floor}\left(\frac{1}{2} + \frac{\omega - K + C}{2K}\right) \tag{91}$$

In the limit when $\kappa_0 \to 0$ the cantilever free end becomes an inflection point and when $C \to 0$ the free end becomes a compression point.

- *Case when $\kappa_0 > 0$*. In this case $K < C \leq 2K$ the number of inflection points *m* and the number of compression points *m'* are given by

$$m = \text{floor}\left(\frac{1}{2} + \frac{\omega - 2K + C}{2K}\right) \qquad m' = 1 + \text{floor}\left(\frac{\omega - 2K + C}{2K}\right) \tag{92}$$

Again, in the limit when $\kappa_0 \to 0$ we have $C = K$ and the free end becomes an inflection point. When $\alpha = 0$ we have $C = 2K$ so the free end point becomes a compression point.

- *Case when* $\kappa_0 = 0$. The number of inflection points $m$ and the number of compression points $m'$ are given by

$$m = 1 + \text{floor}\left(\frac{\omega}{2K}\right) \quad m' = \text{floor}\left(\frac{1}{2} + \frac{\omega}{2K}\right) \tag{93}$$

where a cantilever free end is the first inflection point. From this relation we see that

- o  when $\omega$ is constant and $\alpha$ increases then $K$ also increases and therefore $m$ decreases;
- o  when $\alpha$ is constant and $\omega$ increases then $m$ also increases;
- o  when $m$ is constant then $\omega$ and $\alpha$ cannot be independent.

Torque dominant case. Non-inflectional elastica.

For the torque dominant case the curvature is given by (36). Since function $dn$ has no zeros a cantilever in this case can have no inflection points (see Figure 8 left)

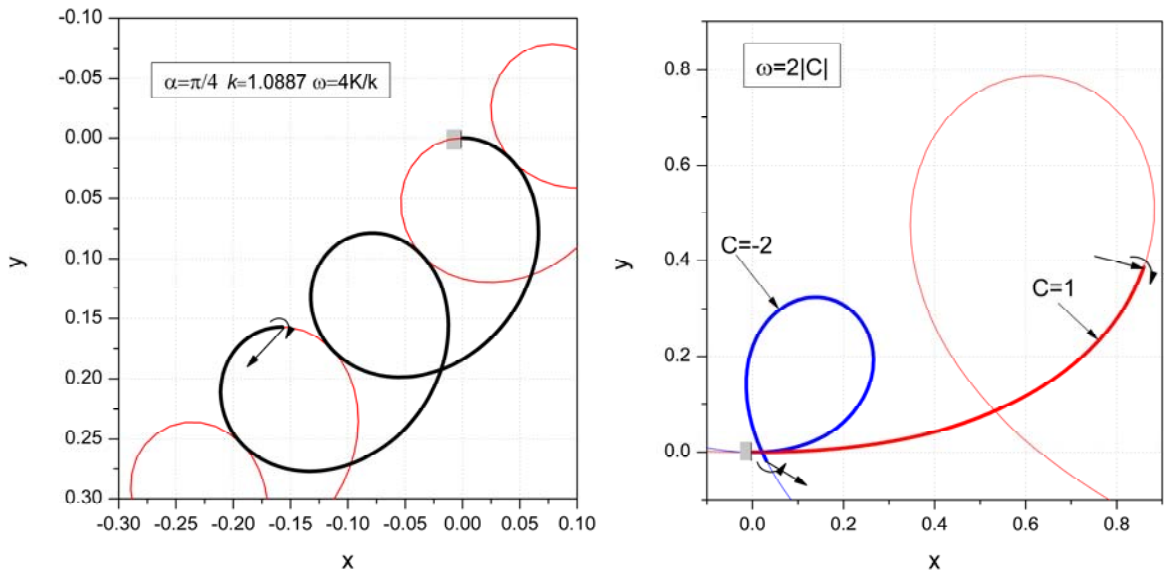

**Figure 8.** The examples of non-inflectional cantilevers. Thin line represents underlying elastica.

*Elastica.* In order to obtain simple forms of equations of elastica we translate the coordinate system into the point where $s_0 = -C/(k\omega)$ and introduce the new parameter coordinate $\sigma$ defined by

$$\sigma = C + k\omega s \tag{94}$$

By this and (56) the parametric equations of elastica are

$$\bar{\xi} = k\omega \left[ \xi \left( s_0 + \frac{\sigma}{k\omega} \right) - \xi(s_0) \right] = \left[ 2k^2 \left( 1 - \frac{E}{K} \right) - 1 \right] \sigma - 2k^2 Z(\sigma) \tag{95}$$

$$\bar{\eta} = k\omega \left[ \eta \left( s_0 + \frac{\sigma}{k\omega} \right) - \eta(s_0) \right] = 2k^2 \left[ 1 - \mathrm{dn}(\sigma) \right]$$

Since both *dn* and *Z* functions have period 2*K* the elastica also has this period. One wave is thus bounded to $0 \le \sigma \le 2K$ and the scale factor is $k\omega$. The tangent angle and the curvature are by (50) and (51)

$$\varphi(s) = -2\mathrm{am}(\sigma, k^{-1}) \qquad \bar{\kappa}(s) \equiv \frac{\kappa(s)}{\omega k} = 2\mathrm{dn}(\sigma, k^{-1}) \tag{96}$$

Similar expressions with different parameterization are given by Goss [10] (Eq 4.26, Eq 4.28, Eq 4.23 and Eq 4.22). By definition at a compression point curvature has extremum. From condition $\frac{d\kappa}{ds} = 0$ by using (51) we obtain the equation $\mathrm{sn}(\sigma, k^{-1}) \mathrm{cn}(\sigma, k^{-1}) = 0$. The parametric coordinates of compression points are zeros of this equation and are at $\sigma'_n = nK$ where *n* is any integer. When $\sigma'_n = 2nK$ we have $\frac{d^2\kappa}{ds^2} = -2\omega^3 k < 0$ and therefore the curvature has maximum

$$\bar{\kappa}_{max} = 2 \tag{97}$$

When $\sigma'_n = (2n+1)K$ we have $\frac{d^2\kappa}{ds^2} = 2\omega^3 \sqrt{k^2 - 1} > 0$ and therefore the curvature has minimum

$$\bar{\kappa}_{min} = 2\sqrt{1 - k^{-2}} \tag{98}$$

All the compression points where curvature has a maximum lie on the line $\bar{\eta} = 2k\left(k - \sqrt{k^2 - 1}\right)$, while all the compression points where curvature is minimal lie on the line $\bar{\eta} = 0$. We note that all the lines of the form $\bar{\eta} = c$ are in space coordinates given by

$$x \sin \gamma + y \cos \gamma = \frac{c}{k\omega} - \eta(s_0) = \frac{c - 2k^2 \left[1 - \mathrm{dn}(k\omega + C, k^{-1})\right]}{k\omega} \tag{99}$$

*Dimensions*. From (95) the distance between wave end points is

$$\Delta \bar{\xi}_c = \bar{\xi}(2K) - \bar{\xi}(0) = 2\left[2k^2(K - E) - K\right] \tag{100}$$

The two end points coincide when $\Delta\bar{\xi}_c = 0$ and this is possible only when $k = \infty$ i.e., in the case with no applied force or when elastica becomes a circle. The wave extreme points in $\bar{\xi}$ direction are obtained from the condition $\dfrac{d\bar{\xi}}{d\sigma} = 2\mathrm{sn}^2(\sigma) - 1 = 0$. At interval $\sigma \in (0, 2K)$ we have two solutions

$$\sigma_1 = \mathrm{sn}^{-1}\left(\frac{\sqrt{2}}{2}\right) \qquad \sigma_2 = 2K - \sigma_1 \tag{101}$$

The distance between two extremes $\Delta\bar{\xi}_t$ in horizontal direction is

$$\Delta\bar{\xi}_t = \bar{\xi}(\sigma_2) - \bar{\xi}(\sigma_1) = 2\left[2k^2\left(\frac{E}{K} - 1\right) + 1\right](1 - \sigma_1) + 4k^2 Z(\sigma_1) \tag{102}$$

and the distance $\Delta\bar{\eta}_t$ between the extreme point and compression point is

$$\Delta\bar{\eta}_t = \bar{\eta}(\sigma_1) = \bar{\eta}(\sigma_2) = k\left(2k - \sqrt{4k^2 - 2}\right) \tag{103}$$

The wave extremes in $\bar{\eta}$ direction are obtained by solution of $\dfrac{d\bar{\eta}}{d\sigma} = 2\mathrm{sn}(\sigma)\mathrm{cn}(\sigma) = 0$. At the interval $\sigma \in [0, K)$ we have two possible values

$$\sigma_1 = 0 \qquad \sigma_2 = K \tag{104}$$

The height of wave $\Delta\bar{\eta}_c$ is therefore

$$\Delta\bar{\eta}_c = \bar{\eta}(K) - \bar{\eta}(0) = 2k\left(k - \sqrt{k^2 - 1}\right) \tag{105}$$

*Self-intersection points.* Similar to the inflectional case we set intersection points that are at $\sigma_1 = \zeta$ and $\sigma_2 = -\zeta + qK$ where $q$ is an integer. By (95) the parameter $\zeta$ is the solution of the following equation

$$\Phi \equiv \left[2k^2\left(\frac{E}{K} - 1\right) + 1\right]\left(\zeta - \frac{qK}{2}\right) + 2k^2 Z(\zeta) = 0 \qquad (0 < \zeta < K) \tag{106}$$

For the case $q = 0$ this equation reduces to

$$\left[2k^2\left(1 - \frac{E}{K}\right) - 1\right]\zeta - 2k^2 Z(\zeta) = 0 \tag{107}$$

At the given interval $Z \geq 0$ and the first term is also positive so we can in any case obtain the intersection point. This can be also observed on the graph in Figure 6 (right). To obtain the case when successive waves touch we solve $\frac{\partial \Phi}{\partial \zeta} = 1 - 2\text{sn}^2(\zeta, k^{-1}) = 0$ which gives $\zeta = \text{sn}^{-1}\left(\frac{\sqrt{2}}{2}, k^{-1}\right)$. For $q = -1$ this by (106) gives the value $k \approx 1.08874$. In this case the intersection point is given by parameter $\zeta \approx 1.82566$. The shape and dimensions of elastica for this case are shown in Figure 7 (case d).

*Cantilever.* The end points of a cantilever in the new parameterization are given by

$$\sigma_0 = C \qquad \sigma_1 = C + k\omega \tag{108}$$

The number of compression points is therefore

$$m' = \text{floor}\left(\frac{k\omega + C}{K}\right) + \begin{cases} 0 & \alpha < 0 \\ 1 & \alpha > 0 \end{cases} \tag{109}$$

When $\alpha > 0$ (torque has the same sense as torque produced by applied force) then origin is always part of cantilever; i.e., cantilever in this case contains at least one compression point.

Case *k*=1. Homoclinic elastica.

*Elastica.* We introduce the new parameter coordinate $\sigma$ defined by

$$\sigma = C + \omega s \tag{110}$$

and translate the coordinate system into the point $s_0 = -C/(k\omega)$. From (64) the parametric equations of elastica are

$$\bar{\xi} \equiv \omega\left[\xi\left(s_0 + \frac{\sigma}{\omega}\right) - \xi(s_0)\right] = \sigma - 2\tanh(\sigma)$$

$$\bar{\eta} \equiv \omega\left[\eta\left(s_0 + \frac{\sigma}{\omega}\right) - \eta(s_0)\right] = 2\left[1 - \frac{1}{\cosh(\sigma)}\right] \tag{111}$$

Both functions are non-periodic so $\sigma \in (-\infty, \infty)$. Again apart from parameterization these formulas are similar to those given by Goss ([10], Eq 4.57, Eq 4.58). The example of homoclinic elastic is shown on Figure 8 (right).

The tangent angle and the curvature are in the new parameterization given by (63)

$$\varphi(s) = -2\sin^{-1}\left[\tanh(\sigma)\right] \qquad \bar{\kappa} \equiv \frac{\kappa(s)}{\omega} = \frac{2}{\cosh(\sigma)} \tag{112}$$

By definition at compression points curvature has extremum. The condition $\frac{d\kappa}{ds}=0$ by using (63)$_2$ gives the equation $\sinh(\sigma)=0$. This equation has only one zero $\sigma'=0$. At this point we have $\frac{d^2\kappa}{ds^2}=-2\omega^3<0$ and therefore the curvature has maximum

$$\kappa_{max}=2 \tag{113}$$

*Dimensions.* From (111) we can derive various dimensions of elastica. The wave extreme points in $\bar{\xi}$ direction are obtained from the condition that at $\frac{d\bar{\xi}}{d\sigma}=1-2/\cosh^2\sigma=0$ and this gives two values

$$\sigma_1=\cosh^{-1}(\sqrt{2})\approx 0.88137 \qquad \sigma_2=-\sigma_1 \tag{114}$$

so the arc length between extremes is

$$\Delta\sigma=\sigma_2-\sigma_1=2\cosh^{-1}(\sqrt{2})\approx 1.76275 \tag{115}$$

The horizontal distance between extremes is

$$\Delta\bar{\xi}_t=\bar{\xi}(\sigma_1)-\bar{\xi}(\sigma_2)=2(\sqrt{2}-\cosh^{-1}\sqrt{2})\approx 1.06568 \tag{116}$$

and the distance between compression point and extreme is

$$\Delta\bar{\eta}_t=\bar{\eta}(\sigma_1)-\bar{\eta}(0)=2-\sqrt{2}\approx 0.58579 \tag{117}$$

In the limit we have

$$\lim_{\sigma\to\pm\infty}[\bar{\eta}(\sigma)-\bar{\eta}(0)]=2 \tag{118}$$

Shape and dimensions of homoclinic elastica are shown in Figure 7 (case c).

*Self-intersection point.* Again we set that intersection points are at $\sigma_1=\zeta$ and $\sigma_2=-\zeta$. By (111) we then obtain the following equation

$$\zeta-2\tanh\zeta=0 \tag{119}$$

which has the solution $\zeta\approx 1.91501$ and gives $\bar{\eta}\approx 1.42316$.

*Cantilever.* The end points of a cantilever are given by

$$\sigma_0=C \text{ and } \sigma_1=C+\omega \tag{120}$$

so the cantilever contains the compression point only if $C \leq 0$ i.e., when $\alpha > 0$. In the special case when $C = -\zeta$ where $\zeta$ is the solution of (119) we by (65) obtain $\alpha \approx 0.81374\pi$ or $\alpha = 146^0 28' 23.7''$. At this angle by (120) a cantilever deforms into a closed loop when $\omega = 2\zeta$.

The results of this section is summarizes on the graph in Figure 9 where a various forms of elastic as a function of *k* and the distance between compression points are shown.

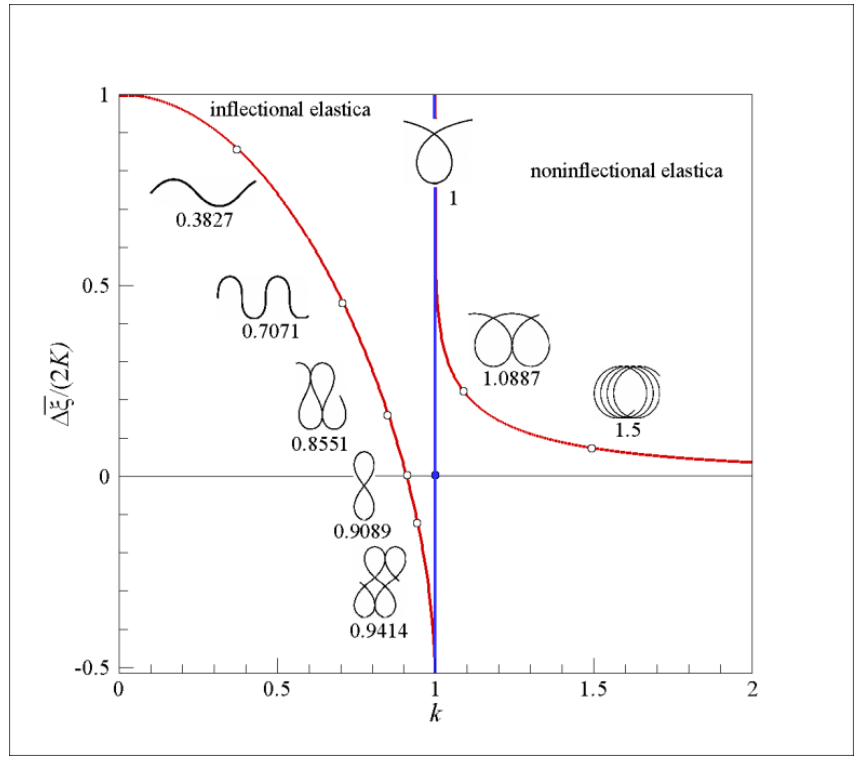

**Figure 9.** Various forms of elastica as a function of *k* and the distance between compression points

**7 Various force load conditions**

In this section we will discuss various problems which can be set by relation (20) where we will discuss only the cases of applied force $\kappa_0 = 0$. In particular we have the following problems:

1. When $\omega$ and $\alpha$ are given then (20) is the explicit expression for $\gamma$ so we have only one solution. This problem is known as the follower load problem. In this case the force is non-conservative since its line of action depends on a deformed cantilever shape.
2. When $\alpha$ and $\gamma$ are given then (20) represents the equation for an unknown $\omega$. Since the equation is in general nonlinear we may expect multiple solutions; that is, multiple equilibrium configurations.

As will be shown, we in fact obtain infinitely many solutions ([44]). We will call this problem a load parameter problem.

3. When $\omega$ and $\gamma$ are given then (20) represents the equation for unknown $\alpha$. This is a conservative load problem and has a finite number of solutions ([33], [35]).

In the end of this section we will also discuss the rotation load problem which is a generalization of the above problems.

<u>Follower load</u> When $\omega$ and $\alpha$ are given we have the follower load problem. In this case Eq (70) when $|\alpha|<\pi$ gives an explicit and unique solution for $\gamma$. Two examples are shown in Figures 10 and 11. In the first of these figures the load parameter increase and $\alpha$ is constant, in second figure the situation is the opposite. We note that all other possibilities of input data yield to multiple solutions of the nonlinear equation (70) that is to finite or infinite number of equilibrium configurations. However, each of them can be reached by some equivalent follower load.

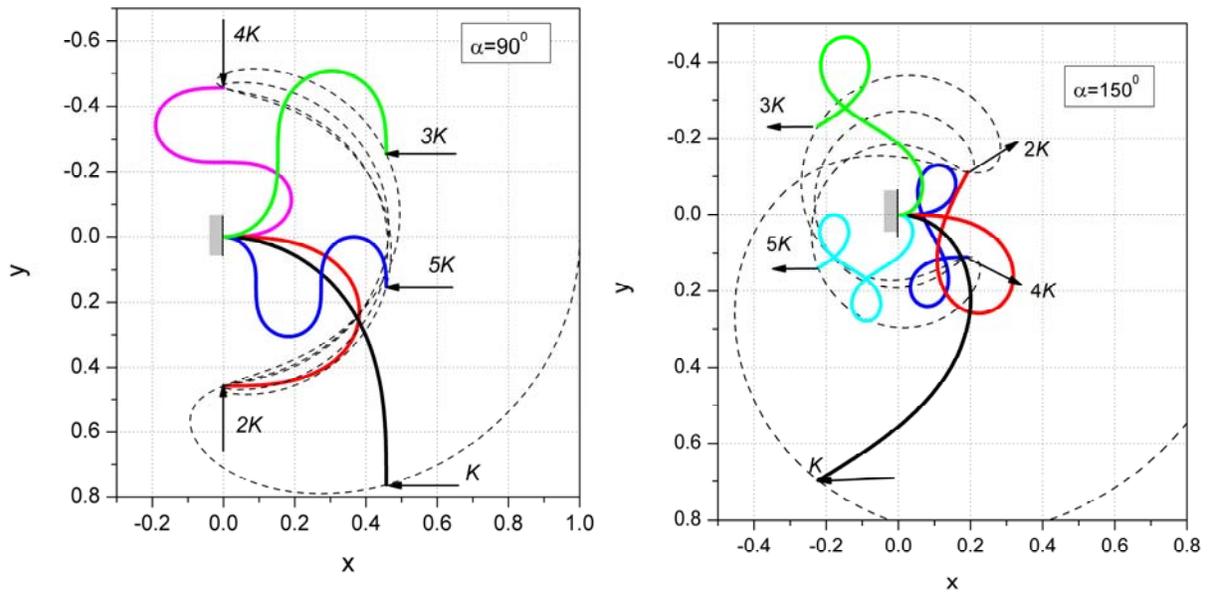

**Figure 10.** Path of beam tip point (dotted line) and some correspondent beam shapes when $\alpha$ is given and $\omega$ increases. Note that with increasing $\omega$ the number of a cantilever's waves also increases.

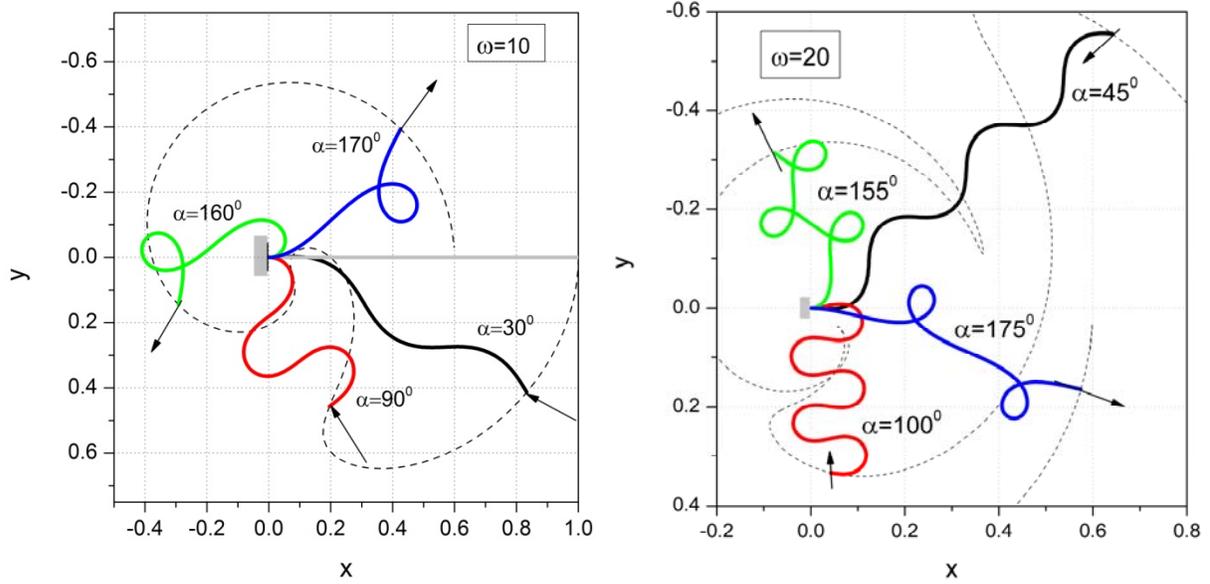

**Figure 11.** The path of beam tip point (dotted line) and some correspondent beam shapes when $\omega$ is given and $\alpha$ increases. Note that with increasing $\alpha$ the number of a cantilever's waves decreases.

We now consider some particular solutions of (70) which follow the form of special values of the Jacobian elliptic functions.

1. When $\omega_n = (2n-1)K$ $(n=1,2,...)$ then the cantilever fixed point is a compression point and from (70) we have

$$\gamma = 0 \qquad (121)$$

   which means that the force acts in the horizontal direction. The free point tangent angle and free point coordinates are

$$\phi_0 = \alpha \qquad x_0 = 2E/K - 1 \qquad y_0 = (-1)^n \frac{2k}{(2n+1)K} \qquad (122)$$

   We see that $x_0$ is independent of a particular load while $y_0$ tends to zero with an increasing load.

2. When $\omega_n = 2(2n-1)K$ $(n=1,2,...)$ the cantilever shape is formed by $n-1$ waves followed by a half wave. Eq (70) in this case gives

$$\gamma = -\alpha \qquad (123)$$

   The cantilever free point tangent angle and coordinates are in this case

$$\phi_0 = 2\alpha \qquad x_0 = (2E/K - 1)\cos\alpha \qquad y_0 = (2E/K - 1)\sin\alpha \qquad (124)$$

Both cantilever free point coordinates are independent of the particular value of the load parameter.

3. When $\omega_n = 4nK$ $(n = 1, 2, \ldots)$ the cantilever shape is formed by *n* waves and we from (70) obtain

$$\gamma = \alpha \qquad (125)$$

Like in the previous case the coordinates of tip points are independent of load factor and are given by

$$\phi_0 = 0 \qquad x_0 = (2E/K - 1)\cos\alpha \qquad y_0 = -(2E/K - 1)\sin\alpha \qquad (126)$$

The described behaviors given by these special solutions may be observed in Figure 10.

*Series expansion for small α.* When α is small then by (66) we have $k \approx \alpha/2$. The power series expansions of (68) and (69) with respect to *k* are

$$\varphi(s) = 2k\cos(\omega s) + O(k^3) \qquad \kappa(s) = 2\omega k \sin(\omega s) + O(k^3) \qquad (127)$$

In particular, we have

$$\gamma = 2k\cos\omega + O(k^2) \qquad (128)$$

so $\cos\gamma \approx 1$ and $\sin\gamma \approx \gamma$. The shape of the beam base curve given by (19) and (44) is therefore approximated by

$$x = (1-s) + O(k^2) \qquad y = -k\left[(1-s)\cos\omega - \frac{\sin\omega - \sin(\omega s)}{\omega}\right] + O(k^3) \qquad (129)$$

By eliminating *s* from these equations we obtain

$$y(x) = \frac{\alpha}{\omega}\left[(1 - \cos\omega x)\sin\omega - (\omega x - \sin(\omega x))\cos\omega\right] + O(\alpha^3) \qquad 0 \leq x \leq 1 \qquad (130)$$

which shows that the deviation of a catilever from a straight line is small. The function (130) satisfies boundary conditions $y(0) = 0$, $y'(0) = 0$ and $y''(1) = 0$ while $y'(1) = \alpha(\sin\omega/\omega - \cos\omega)$. We note that this form of solution was obtained by Pflüger ([36],pp 217-218) in the study of the elastic stability of the load follower cantilever beam. Also we note that the solution given by Saalschütz ([3],pp) does not satisfy boundary condition $y''(1) = 0$. Table 6 contains some comparisons of results of calculations obtained by exact and approximate solutions.

**Table 6.** Comparison of result of exact and approximate calculation of cantilever free end coordinates $(x_0, y_0)$, free end tangent angle $\phi_0$ and root curvature $\kappa_1$ for given $k$ and $\omega$

| method | $k$ | $\alpha$ (deg) | $\omega$ | $x_0$ | $y_0$ | $\phi_0/\pi$ | $\kappa_1/\omega$ |
|---|---|---|---|---|---|---|---|
| exact |  |  |  | 0.9769 | 0.1579 | 0.1181 | -0.1043 |
| approx. | 0.1 | 11.5 | 10 | 1.0000 | 0.1569 | 0.1170 | -0.1088 |
| rerr % |  |  |  | 2.36 | 0.63 | 0.90 | 4.25 |

*Small load parameter.* When $\omega$ is small then we have the following expansions

$$\varphi(s) = \alpha - \sin\alpha \frac{\omega^2 s^2}{2} + O(\omega^4) \tag{131}$$

$$\kappa(s) = \omega^2 \sin\alpha \, s + O(\omega^4) \tag{132}$$

$$\xi(s) = (1-s)\left[\cos\alpha + \frac{\sin^2\alpha}{6}(1+s+s^2)\omega^2\right] + O(\omega^4)$$

$$\eta(s) = (1-s)\left[\sin\alpha - \frac{\sin(2\alpha)}{12}(1+s+s^2)\omega^2\right] + O(\omega^4) \tag{133}$$

In the limit case $\omega \to 0$ we have

$$\xi(s) \to (1-s)\cos\alpha \quad \eta(s) \to (1-s)\sin\alpha \quad \gamma \to \alpha \tag{134}$$

and this by (19) approaches trivial solution (23).

*Large load parameter.* When $\omega$ is large we from (71) obtain the following asymptotic estimations

$$\xi(s) = (2E/K - 1)(1-s) + O(\omega^{-1}) \quad \eta(s) = O(\omega^{-1}) \tag{135}$$

and therefore we have by (19)

$$x(s) = \xi(s)\cos\gamma + O(\omega^{-1}) \quad y(s) = -\xi(s)\sin\gamma + O(\omega^{-1}) \tag{136}$$

This shows that when $\omega$ is large the shape of the cantilever when $\gamma \neq \pi/2$ approaches line $y = -\tan\gamma \, x$ and when $\gamma = \pi/2$ the line $x = 0$. In other words, with increasing load the beam becomes more and more straight. This can be observed in Figure 13 where two cases are present.

The local coordinates of the cantilever free point are by (135)

$$\xi(0) = 2E/K - 1 + O(\omega^{-1}) \qquad \eta(0) = O(\omega^{-1}) \tag{137}$$

and its space coordinates are by (136)

$$x(0) = (2E/K - 1)\cos\gamma + O(\omega^{-1}) \qquad y(0) = -(2E/K - 1)\sin\gamma + O(\omega^{-1}) \tag{138}$$

When $\omega$ is large and $\alpha$ is constant then the path described by the cantilever free point becomes approximately a circular arc. The arc are for $2E > K$ bounded to the interval $-\alpha \leq \gamma \leq \alpha$ and for $2E < K$ to the interval $|\gamma| \leq \pi - \alpha$. This may be seen in Figure 10.

*Load parameter problem.* In the case when $\alpha$ and $|\gamma| \leq \alpha$ are given (70) becomes the equation for an unknown $\omega$. We rewrite this equation into the form

$$\operatorname{sn}(\omega + K) = A \qquad A \equiv \frac{\sin(\gamma/2)}{\sin(\alpha/2)} \tag{139}$$

We note that this problem also covers the problem when the free point tangent angle $\phi_0$ and $\alpha$ are given since then $\gamma$ is given by (5).

The solution of this equation that is closest to the origin is

$$\omega_0 = K - \operatorname{sn}^{-1}(A, k) \tag{140}$$

where $0 \leq \omega_0 \leq K$. From periodicity of function *sn* then follows the infinite sequence of possible solutions. We distinguish two cases:

1. when $\gamma \geq 0$ we have the sequence of load parameters

$$\omega_{2n-1} = \omega_0 + 4(n-1)K \qquad \omega_{2n} = -\omega_0 + 4nK \qquad (n = 1, 2, \ldots) \tag{141}$$

2. and when $\gamma < 0$ the load parameters are

$$\omega_{2n-1} = -\omega_0 + 2(2n-1)K \qquad \omega_{2n} = \omega_0 + 2(2n+1)K \qquad (n = 1, 2, \ldots) \tag{142}$$

Load parameters given by (141) and (142) can be represented as branches on a bifurcation diagram in a $(\omega, \alpha)$ plane as is shown on graph in Figure 12. Note that there are two possible uses of solution (141) or (143). If we take the constant *n* (which is also called the wave number because it determines the number of waves that forms the cantilever shape) then the solution of the problem is a load parameter that is a continuous function of $\alpha$. When $\alpha$ is constant then we for each *n* obtain an equilibrium configuration. The examples are shown in Figure 13 and Figure 14.

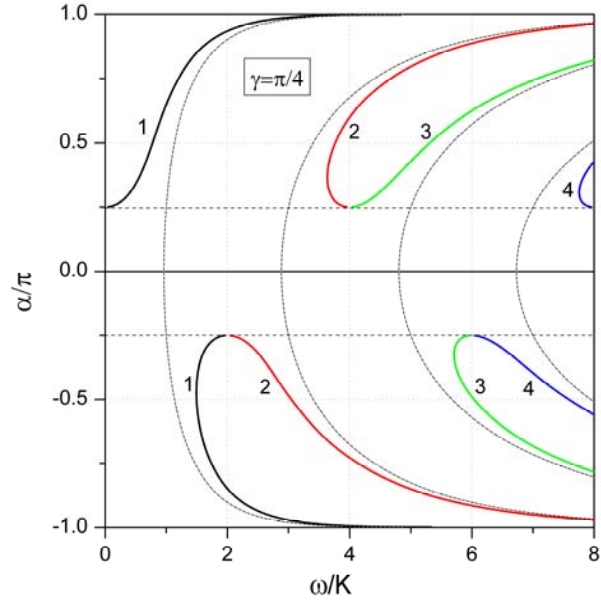

**Figure 12.** Graph of (141) for $\gamma = \pi/4$. Dot curve represents the graph for case $\gamma = 0$.

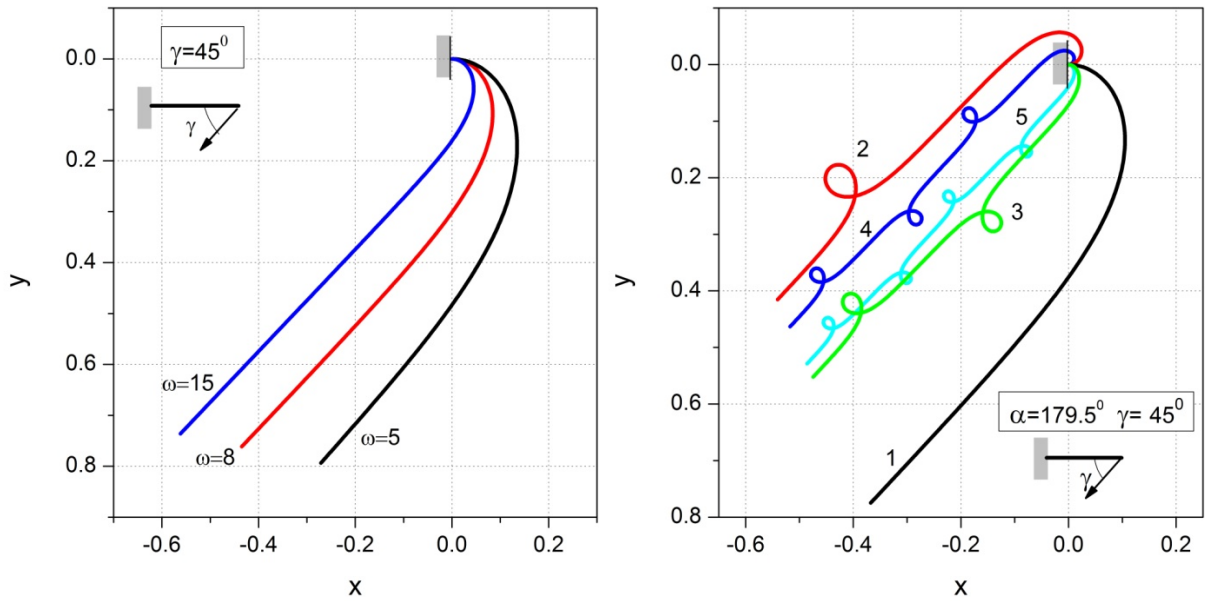

**Figure 13.** Cantilever beam equilibrium configurations when $\gamma = 45^0$ calculated by (141). In the left figure $\alpha$ is calculated for various load parameters $\omega$ when $n=1$. In the right figure successive load parameters $\omega$ are calculated for a given $\alpha = 179.5^0$ and various $n$. In both cases a cantilever with increasing $\omega$ becomes more and more straight.

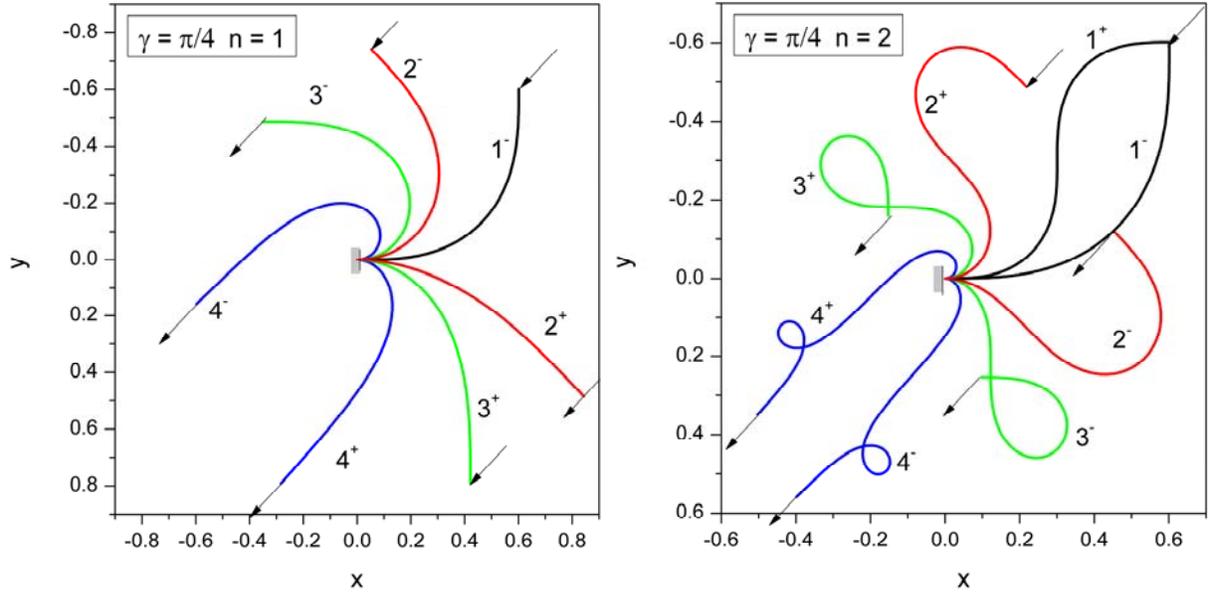

**Figure 14**. Equilibrium shapes when $\gamma = \pi/4$, $\alpha/\pi = (\pm 0.25, \pm 0.5, \pm 0.75, \pm 0.99)$ and different *n*. Load parameters are calculated by (141). Successive numbers correspond to successive values of $\alpha$.

<u>Conservative load</u> Consider now the case when $\gamma$ and $\omega$ are given and $\alpha$ is unknown. This is the problem of conservative load. We note that when $\gamma$ and $\phi_0$ are given then $\alpha = \gamma + \phi_0$ and the problem is equivalent to the load parameter problem that was discussed in the previous section.

We rewrite (70) into the following form

$$\sin\frac{\gamma}{2} = k\,\mathrm{sn}(\omega + K, k) \qquad \alpha = 2\sin^{-1}(k) \tag{144}$$

and consider two cases of possible solutions for unknown *k* where we assume that $k \neq 0$.

*The case when $\gamma = 0$.* In this case the solutions of (144) are $\omega = (2n-1)K$ $(n = 1, 2, \ldots)$. For a given $\omega$ we therefore have the equation

$$K(k) = \frac{\omega}{(2n-1)} \qquad (n = 1, 2, \ldots) \tag{145}$$

and this, because $K(k)$ is a monotone function, for each *n* gives a unique value of *k*. However, since $K(k) \geq \pi/2$ and the right hand side of (145) tends to zero with increasing *n* the number of solutions is finite. In order to determine the number of solutions we first consider the special case when $\alpha = 0$. In this case $K(0) = \pi/2$ and therefore $\omega_n = (2n-1)\frac{\pi}{2}$ $(n = 1, 2, \ldots)$. These values of $\omega$ represent bifurcation points on $(\omega,\alpha)$ plane (see Figure 12). We see that the number of possible equilibrium configurations doubles at each bifurcation point. So if $\omega_n \leq \omega < \omega_{n+1}$ then the number of possible equilibrium configurations including the trivial solution is $2n+1$ where *n* is

$$n = \text{floor}\left(\frac{\omega}{\pi}+\frac{1}{2}\right) \tag{146}$$

For example for $\omega=15$ we have $n=5$ and therefore 11 possible equilibrium configurations.

Equation (145) does not have an analytical solution and must be solved numerically. The initial estimation of the solution is obtained on the basis of inequality. $\ln 4 \leq K + \ln k' \leq \pi/2$ ([64],pp 318) from which it follows that the *m*-th zero falls into the interval

$$\sqrt{1-e^{\pi}e^{-2\omega/(2m-1)}} \leq k_m \leq \sqrt{1-16e^{-2\omega/(2m-1)}} \qquad (m=1,2,\ldots,n) \tag{147}$$

Once we obtain $k_m$ we have $\alpha_m = 2\sin^{-1}(k_m)$. Zeros for $\alpha<0$ are then obtained by symmetry.

*Case* $0<\gamma<\pi$. When $0<\gamma<\pi$ then (144) must be solved numerically where roots are to be located in the interval

$$-\pi < \alpha < \pi$$

From the shape of the function (70) shown in Figure 15 (right) we see that possible roots lie within intervals bounded by roots of (145) and that we have two possible roots within intervals. These observations suggest that for calculation of roots of (144) we may use the following procedure:

1. Calculate number of zeros by (146)
2. Calculate zeros by using estimation (147)
3. Calculate location of extreme points between two neighboring zeros and add to them extreme values which are on the end of the interval $-\pi<\alpha<\pi$
4. Make partition of the interval $-\pi<\alpha<\pi$ to subintervals where each subinterval is bounded by extreme and zero point
5. For each such interval calculate possible zero of (144)

The example of calculation by this procedure is shown in Figure 15 and some numerical values are given in Table 7.

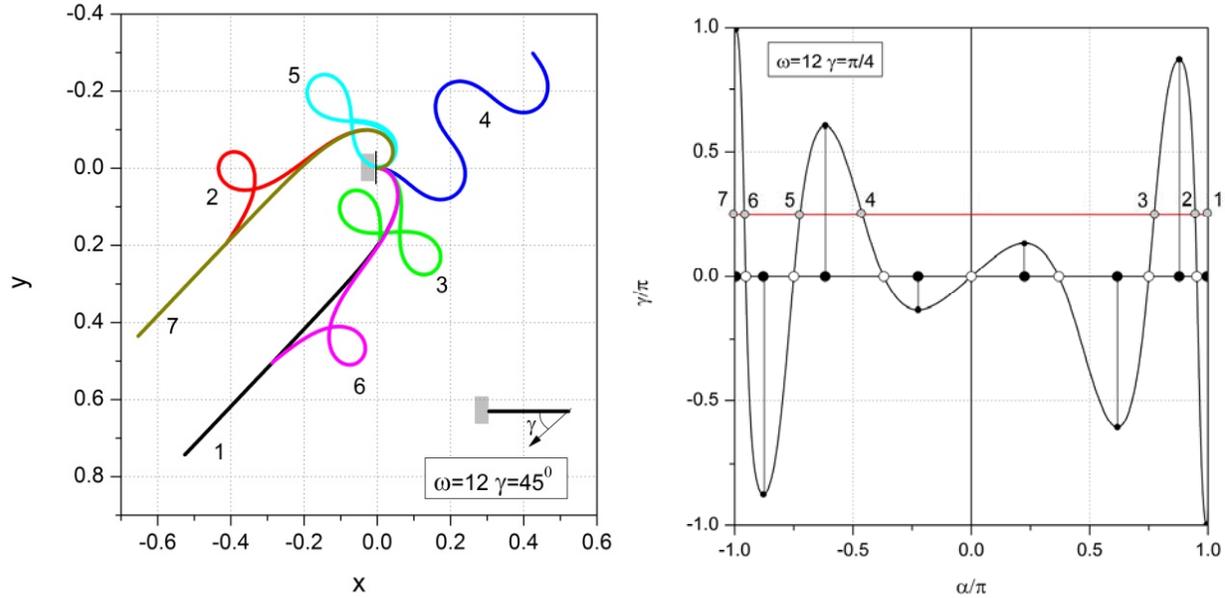

**Figure 15.** The cantilever equilibrium shapes when $\omega=12$ (left). Partition of the interval when $\omega=12$ (right). For this case there are 18 intervals, however the two near each ends are very narrow and cannot be displayed well. The number of possible equilibrium configurations is seven.

**Table 7.** Calculated cantilever free end coordinates $(x_0, y_0)$, free end tangent angle $\phi_0$ and root curvature $\kappa_1$ for equilibrium configurations for the case of conservative load when $\gamma = 45^0$ and $\omega=12$. Numbers correspond to shapes in Figure 15.

|   | $\alpha/\pi$ | $x_0$ | $y_0$ | $\phi_0/\pi$ | $\kappa_1/\omega$ | $\gamma/\pi$ |
|---|---|---|---|---|---|---|
| 1 | 0.999989546 | -0.525475296 | 0.743235391 | 0.749989649 | 1.847759190 | 0.249999897 |
| 2 | 0.946286875 | -0.412482787 | 0.195630101 | 0.696286875 | -1.840056050 | 0.250000000 |
| 3 | 0.773219245 | 0.006563252 | 0.195087408 | 0.523219245 | 1.711062590 | 0.250000000 |
| 4 | -0.460401429 | 0.425240149 | -0.297980044 | -0.710401429 | 1.079837790 | 0.250000000 |
| 5 | -0.722781359 | -0.068424812 | -0.125313200 | -0.972781359 | -1.643921540 | 0.250000000 |
| 6 | -0.959076697 | -0.287482201 | 0.504715675 | -1.209076700 | 1.843287150 | 0.250000003 |
| 7 | -0.999976584 | -0.653036197 | 0.435275758 | -1.249976600 | -1.847759040 | 0.250000016 |

Rotational load. The rotational load generalizes the pure tip follower load into the following form

$$\alpha = \alpha_0 + (1-\beta)\phi_0 \qquad 0 \le \alpha_0 \le \pi \qquad 0 \le \beta \le 1 \tag{148}$$

where $\alpha_0$ is a constant angle and $\beta$ is the rotation factor. For $\beta=1$ we have a pure follower load and for $\beta=0$ we have a pure conservative load. This type of load is a bit artificial and was for the special case $\alpha_0 = \pi/2$ introduced by Nageswara and Venkateswara ([44]).

Now, by using (148) and (5)$_1$ and identity $sn(\omega+K)=cn(\omega)/dn(\omega)$ we from (70) obtain the governing equation of the problem in the following form

$$\sin\frac{\alpha_0-\beta\phi_0}{2}dn(\omega,k)=k\,cn(\omega,k) \qquad k=\sin\frac{\alpha_0+(1-\beta)\phi_0}{2} \tag{149}$$

We will here, opposite to [44, 65], consider the case when in the above equation the cantilever free point tangent angle $\phi_0$ is unknown and load parameter $\omega$ and rotation factor $\beta$ are given. Namely, when $\alpha_0$, $\beta$ and $\phi_0$ are given then we can by (148) calculate in advance $\alpha$ and future by (5)$_1$ also $\gamma=\alpha_0-\beta\phi_0$. Therefore, the problem is in this case equivalent to the load parameter problem.

Some special solutions (149) are:

1. When $\omega_n=4nK$ then $cn(\omega_n,k)=1$ and $dn(\omega_n,k)=1$. Consequently $\alpha_0-\beta\phi_0=\alpha_0+(1-\beta)\phi_0$ so $\phi_0=0$ and future $\gamma=\alpha_0$.

2. When $\omega_n=(2n+1)K$ then $cn(\omega_n,k)=0$ and $dn(\omega_n,k)=\sqrt{1-k^2}\neq 0$. Consequently $\alpha_0=\beta\phi_0$ and $\gamma=0$.

3. When $\omega_n=2(2n+1)K$ then $cn(\omega_n,k)=-1$ and $dn(\omega_n,k)=1$. Consequently $\phi_0=\dfrac{2\alpha_0}{2\beta-1}$ and $\gamma=-\dfrac{\alpha_0}{2\beta-1}$.

In general (149) must be solved numerically. The first step in numerical calculation is determination of the range of $\phi_0$. The first estimation follows from condition $k^2\leq 1$

$$-\frac{\pi+\alpha_0}{1-\beta}\leq\phi_0\leq\frac{\pi-\alpha_0}{1-\beta} \quad (\beta\neq 1) \tag{150}$$

Next, consider the first integral (28) which becomes by using (148)

$$\kappa^2=2\omega^2\left[\cos(\phi+\alpha_0-\beta\phi_0)-\cos(\alpha_0+(1-\beta)\phi_0)\right] \tag{151}$$

To obtain the real solution the right hand side (RHS) of this equation must be non-negative. For a given $\alpha_0$ and $\beta$ this requirement divides the entire $(\phi_0,\phi)$ plane into a network of alternating positively and negatively signed parallelograms. The positively signed parallelograms are those where the solution is possible (white regions in Figure 16). For $\beta=0$ the paralegals are squares and with increasing $\beta$ they continuously deform into parallel strips for $\beta=1$. The equations of lines that bound parallelograms and where the value of the RHS of (151) is zero are

$$\phi=\phi_0+2\pi m \qquad \phi=-(1-2\beta)\phi_0+2(n\pi-\alpha_0) \qquad m,n=0,\pm 1,\pm 2,... \tag{152}$$

Now, the possible integration paths in $(\phi_0, \phi)$ plane that solves the problem are vertical lines which must, in order to satisfy boundary conditions, start at the intersection with the line $\phi = \phi_0$ and end on the line $\phi = 0$ and in addition they must not pass the negatively signed parallelograms. By this observation we can - with the help of the graph shown in Figure 16 – find that the possible solution ranges are governed by the intersection of $m=0$ and $n=1$ lines and $m=1$ and $n=0$ lines and the intersection of $n=0$ line with line $\phi=0$. Detailed analysis shows that depending on the values of rotation parameter $\beta$ there are three possibilities:

1. if $0 \leq \beta < \dfrac{1 - \alpha_0/\pi}{2}$ then $-\dfrac{\pi + \alpha_0}{1 - \beta} < \phi_0 \leq -\dfrac{2\alpha_0}{1 - 2\beta}$ or $0 \leq \phi_0 < \dfrac{\pi - \alpha_0}{1 - \beta}$

2. if $\dfrac{1 - \alpha_0/\pi}{2} \leq \beta < \dfrac{1 + \alpha_0/\pi}{2}$ then $0 \leq \phi_0 < \dfrac{\pi - \alpha_0}{1 - \beta}$

3. if $\dfrac{1 + \alpha_0/\pi}{2} \leq \beta < 1$ then $0 \leq \phi_0 \leq \dfrac{2\alpha_0}{2\beta - 1}$

For $\phi_0 \geq 0$ the range of tangent angle along the beam is

$$-(1 - 2\beta)\phi_0 - 2\alpha_0 \leq \phi \leq \phi_0 \qquad (153)$$

and for $\phi_0 < 0$ it is

$$\phi_0 \leq \phi \leq -(1 - 2\beta)\phi_0 - 2\alpha_0 \qquad (154)$$

Note also that the requirement that $0 \leq \beta < 1$ implies $0 \leq \alpha_0 < \pi$. The range of $\phi_0$ as function of $\beta$ is shown in Figure 16 (right). A result of numerical calculation for the case $\alpha = 90^0$, $\beta = 0.8$ and $\omega = 14.5$ are given in Table 8 and Figure 17 where a bifurcation diagram and correspondent equilibrium configurations of the cantilever are shown.

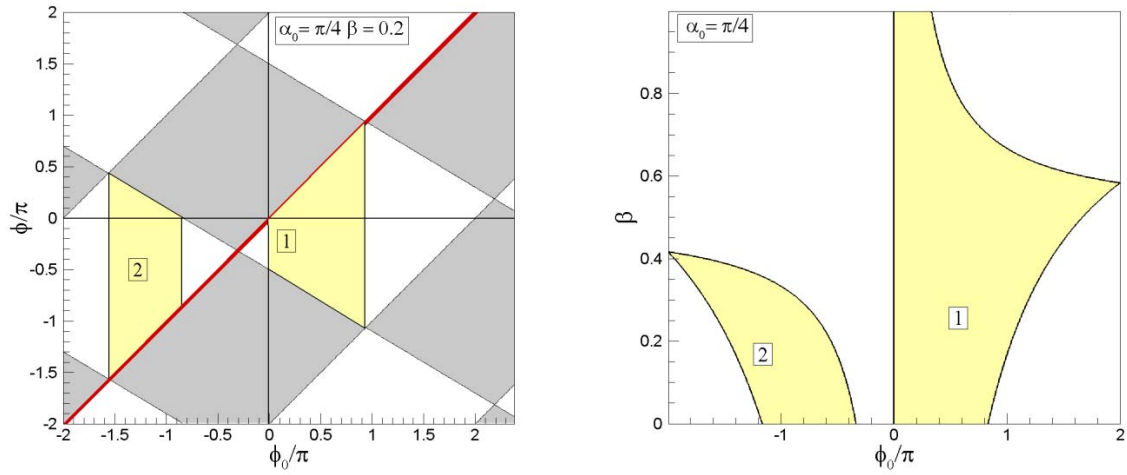

**Figure 16.** Solution regions 1 and 2 on $(\phi_0, \phi)$ plane and distribution of tangent angle along cantilever beam for rotation load when $\alpha_0 = \pi/4$ and $\beta = 0.8$ (left) and solution regions 1 and 2 on $(\phi_0, \beta)$ plane when $\alpha_0 = \pi/4$ (right)

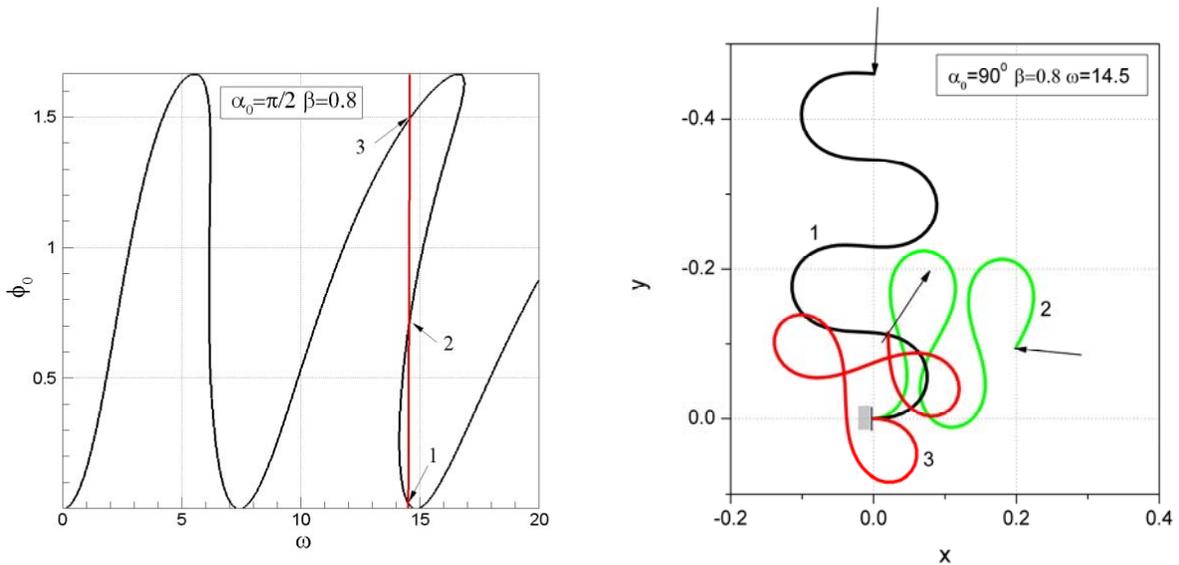

**Figure 17.** Bifurcation diagram for the rotation load when $\alpha_0 = 90^0$ and $\beta = 0.8$ (left). Corresponding cantilever beam equilibrium configurations when $\omega = 14.5$ (right).

**Table 8.** Calculated cantilever free end coordinates $(x_0, y_0)$, free end tangent angle $\phi_0$ and root curvature $\kappa_1$ for equilibrium configurations in the case of rotation load when $\alpha = 90^0$, $\beta = 0.8$ and $\omega = 14.5$ Numbers correspond to shapes in Figure 17.

|   | $x_0$ | $y_0$ | $\phi_0/\pi$ | $\kappa_1/\omega$ | $\alpha/\pi$ | $\gamma/\pi$ |
|---|---|---|---|---|---|---|
| 1 | 0.00056187 | -0.46040271 | 0.02329480 | -0.38248868 | 0.50465896 | 0.48136416 |
| 2 | 0.19934467 | -0.09514761 | 0.66664422 | -1.67406783 | 0.63332884 | -0.03331538 |
| 3 | 0.02005727 | -0.11492198 | 1.48750365 | 0.69611210 | 0.79750073 | -0.69000292 |

**8 Conclusion**

Although the problem discussed in this paper is old and most of the given results are well known, we present some novelties:

1. A new analytical solution of the problem is given in terms of Jacobi elliptical functions and the Jacobi zeta function where cantilever load includes also torque;
2. A new analytical expression for calculating various dimensions of elastica are given
3. We provide a new efficient procedure for determination of all possible equilibrium shapes in the case of the conservative load problem
4. The rotational load problem is defined in a general form and the analysis of the possible domain of tip angle is given.

At the end we add that in the present paper the follower load problem, the load parameter problem, the conservative load problem and the rotation load problem are treated from a single point of view, namely as solutions of equation (70). The Maple worksheet that implements the solution of these problems is freely available at [66]

**References**

1.      Euler, L., *Elastic Curves.* Isis, 1933. **20**(1): p. 72-160.
2.      Clebsch, A., *Theorie der elasticität faster körper*. 1862, Leipzig: B.G.Teubner.
3.      Saalschütz, L., *Der belastete Stab unter Einwirkung einer seitlichen Kraft*. 1880: B. G. Teubner, Leipzig.
4.      Hess, W., *Uber die Biegung und Drillung eines unendlich dünnen elastichen Stabes mit zwei gleichen Widerstanden, auf dessen freies Ende eine Kraft und ein um die Hauptaxe ungleichen Widerstandes drehendes Kräftepaar einwirkt.* Mathematische Annalen, 1885. **25**: p. 1-38.
5.      Love, A.E.H., *A treatise on the mathematical theory of elasticity, Vol. 2*. 2th ed. 1893, London: Cambridge university press
6.      Born, M., *Untersuchungen über die Stabilität der elastischen Linie in Ebene und Raum: unter verschiedenen Grenzbedingungen*. 1906: Ph.D. thesis, University of Gottingen.
7.      Todhunter, I. and K. Pearson, *A history of the theory of elasticity and of the strength of materials : from Galilei to Lord Kelvin*. 1960, New York: Dover Publications. 2 v in 3.

8.  Timoshenko, S., *History of strength of materials : with a brief account of the history of theory of elasticity and theory of structures*. 1953, New York: McGraw-Hill. 452 p.
9.  Truesdell, C. and L. Euler, *The rational mechanics of flexible or elastic bodies, 1638-1788 : introduction to Leonhardi Euleri opera omnia vol X et XI seriei secundae*. Leonhardi Euleri opera omnia. Series 2, Opera mechanica et astronomica. 1960, Turici: Orell F©*ssli. 435 p.
10. Goss, V.G.A., *Snap buckling, writhing and loop formation in twisted rods*, in *Center for Nonlinear Dynamics*. 2003, University Collage London.
11. Levien, R., *The elastica: a mathematical history*. 2008, University of california at Berkeley.
12. Malkin, I., *Formaenderung eines axial gedrueckten duennen Stabes.* Z. angew. Math. Mech., 1926. **6**(73): p. 73-76.
13. Hummel, F.H. and W.B. Morton, *On the Large Bending of Thin Flexible Strips and the Measurment of their Elasticity.* 1924: p. 348-357.
14. Barten, H.J., *On the Deflection of a Centilever Beam.* Quarterly of Applied Mathematics, 1944. **2**: p. 168-171.
15. Bisshopp, K.E. and D.C. Drucker, *Large Deflection of Cantilever Beams.* Quarterly of Applied Mathematics, 1945. **3**(3): p. 272-1945.
16. Timoshenko, S. and J.M. Gere, *Theory of elastic stability*. 2nd ed. Engineering societies monographs. 1961, New York ; London: McGraw-Hill Book Company. 541 p.
17. Mitchell, T.P., *The Nonlinear Bending of Thin Rods.* Transactions of ASME, Journall of Applied Mechanics, 1959: p. 40-43.
18. Beth, R.A. and C.P. Wells, *Finite Deflections of a Cantilever-Strut.* Journal of Applied Physics, 1951. **22**(6): p. 742-746.
19. Scott, E.J., D.R. Carver, and Manhattan, *On the Nonlinear Differential Equation for Beam Deflection.* Transactions of ASME, Journall of Applied Mechanics, 1955. **22**(77): p. 245-248.
20. Frisch-Fay, R., *A New Approach to the Analysis of the Deflection of Thin Cantilevers.* Journal of Applied Mechanics, Transactions of ASME, 1961. **28**: p. 87-90.
21. Frisch-Fay, R., *Flexible bars*. Butterworths scientific publications. 1962, London: Butterworths. viii, 220 p.
22. Massoud, M.F., *On the Problem of Large Deflexion of Cantilever Beam.* Internationa Journal of Mechanical Science, 1966. **8**: p. 141-143.
23. Schmidt, R. and D.A. DaDeppo, *A Survey of Literature on Large Deflections of Nonshallow Arches. Bibliography of Finite Deflections of Straight and Curved Beams, Rings, and Shallow Arches.* Journal of the Industrial Mathematics Society, 1971. **21**(2): p. 91-144.
24. Wang, C.Y., *Large Deflections of an Inclined Cantilever with an End Load.* International Journal of Non-Linear Mechanics, 1981. **16**(2): p. 155-164.
25. Carlson, B.C. and E.M. Notis, *Algorithm 577 - Algorithms for Incomplete Elliptic Integrals [S21].* Acm Transactions on Mathematical Software, 1981. **7**(3): p. 398-403.
26. Mattiasson, K., *Numerical Results from Large Deflection Beam and Frame Problems Analyzed by Means of Elliptic Integrals.* International Journal for Numerical Methods in Engineering, 1981. **17**(1): p. 145-153.
27. Lau, J.H., *Large Deflections of Beams with Combined Loads.* Journal of the Engineering Mechanics Division-Asce, 1982. **108**(1): p. 180-185.
28. DeBona, F. and S. Zelenika, *A generalized elastica-type approach to the analysis of large displacements of spring-strips.* Proceedings of the Institution of Mechanical Engineers Part C-Journal of Mechanical Engineering Science, 1997. **211**(7): p. 509-517.
29. Howell, L.L. and A. Midha, *A Method for the Design of Compliant Mechanisms with Small-Length Flexural Pivots.* Journal of Mechanical Design, 1994. **116**(1): p. 280-290.

30. Saxena, A. and S.N. Kramer, *A simple and accurate method for determining large deflections in compliant mechanisms subjected to end forces and moments.* Journal of Mechanical Design, 1998. **120**(3): p. 392-400.
31. Yau, J.D., *Closed-Form Solutions of Large Deflection for a Guyed Cantilever Column Pulled by an Inclination Cable.* Journal of Marine Science and Technology-Taiwan, 2010. **18**(1): p. 130-136.
32. Popov, E.P., *Theory and Calculation of Flexible Elastic Bars* 1986, Moscow: Nauka.
33. Navaee, S., *Equilibrium-Configurations of Cantilever Beams Subjected to Inclined End Loads - Closure.* Journal of Applied Mechanics-Transactions of the Asme, 1993. **60**(2): p. 564-564.
34. Navaee, S. and R.E. Elling, *Possible Ranges of End Slope for Cantilever Beams.* Journal of Engineering Mechanics-Asce, 1993. **119**(3): p. 630-637.
35. Batista, M. and F. Kosel, *Cantilever beam equilibrium configurations.* International Journal of Solids and Structures, 2005. **42**(16-17): p. 4663-4672.
36. Pflüger, A., *Stabilitätsprobleme der Elastostatik*. 1950, Berlin / Göttingen / Heidelberg: Springer-Verlag.
37. Bolotin, V.V., G. Herrmann, and T.K. Lusher, *Nonconservative problems of the theory of elastic stability*. Corr. and authorized ed ed. 1963, Oxford: Macmillan. xii, 324 p.
38. Argyris, J.H. and S. Symeonidis, *Non-Linear Finite-Element Analysis of Elastic-Systems under Non-Conservative Loading Natural Formulation .1. Quasistatic Problems.* Computer Methods in Applied Mechanics and Engineering, 1981. **26**(1): p. 75-123.
39. Alliney, S. and A. Tralli, *Extended Variational Formulations and Fe Models for Nonlinear Beams under Nonconservative Loading.* Computer Methods in Applied Mechanics and Engineering, 1984. **46**(2): p. 177-194.
40. Saje, M. and S. Srpcic, *Large Deformations of Inplane Beam.* International Journal of Solids and Structures, 1985. **21**(12): p. 1181-1195.
41. Shvartsman, B.S., *Large deflections of a cantilever beam subjected to a follower force.* Journal of Sound and Vibration, 2007. **304**(3-5): p. 969-973.
42. Nallathambi, A.K., C.L. Rao, and S.M. Srinivasan, *Large deflection of constant curvature cantilever beam under follower load.* International Journal of Mechanical Sciences, 2010. **52**(3): p. 440-445.
43. Mutyalarao, M., D. Bharathi, and B.N. Rao, *Large deflections of a cantilever beam under an inclined end load.* Applied Mathematics and Computation, 2010. **217**(7): p. 3607-3613.
44. Nageswara, R.B. and R.G. Venkateswara, *On the Large Deflection of Cantilever Beams with End Rotational Load.* Zeitschrift Fur Angewandte Mathematik Und Mechanik, 1986. **66**(10): p. 507-509.
45. Nageswara, R.B. and R.G. Venkateswara, *Large Deflections of Cantilever Beam Subject to a Tip Concentrated Rotational Load.* Aeronautical Journal, 1986: p. 262-266.
46. Zakharov, Y.V. and A.A. Zakharenko, *Dynamic instability in the nonlinear problem of a cantilever* Vychysl. Tekhnol. (in Russian), 1999. **4**(1): p. 48-54.
47. Zakharov, Y.V. and K.G. Okhotkin, *Nonlinear Bending Of Thin Elastic Rods.* Journal of Applied Mechanics and Technical Physics, 2002. **43**(5): p. 739-744.
48. Zakharov, Y.V., K.G. Okhotkin, and A.D. Skorobogatov, *Bending of Bars Under a Follower Load.* Journal of Applied Mechanics and Tehnical Phyisics, 2004. **45**(5): p. 756-763.
49. Kuznetsov, V.V. and S.V. Levyakov, *Complete solution of the stability problem for elastica of Euler's column.* International Journal of Non-Linear Mechanics, 2002. **37**(6): p. 1003-1009.
50. Levyakov, S.V. and V.V. Kuznetsov, *Stability analysis of planar equilibrium configurations of elastic rods subjected to end loads.* Acta Mechanica, 2010. **211**(1-2): p. 73-87.
51. Wang, J., J.K. Chen, and S.J. Liao, *An explicit solution of the large deformation of a cantilever beam under point load at the free tip.* Journal of Computational and Applied Mathematics, 2008. **212**(2): p. 320-330.

52. Kimiaeifar, A., et al., *Analytical Solution for Large Deflections of a Cantilever Beam Under Nonconservative Load Based on Homotopy Analysis Method.* Numerical Methods for Partial Differential Equations, 2011. **27**(3): p. 541-553.
53. Tari, H., *On the parametric large deflection study of Euler–Bernoullicantilever beams subjected to combined tip point loading.* International JournalofNon-LinearMechanics, 2012. **Article in Press**.
54. Antman, S.S., *Nonlinear problems of elasticity*. 1995, New York ; London: Springer-Verlag. xvii, 750 p.
55. Hirsch, M.W., S. Smale, and R.L. Devaney, *Differential equations, dynamical systems, and an introduction to chaos*. 2nd ed. Pure and applied mathematics a series of monographs and textbooks. 2004, San Diego, Calif. ; London: Academic Press. xiv, 417 p.
56. Greenhill, G., *The applications of elliptic functions*. 1892, London: Macmillan. xi, 357 p.
57. Armitage, J.V. and W.F. Eberlein, *Elliptic functions*. 2006, Cambridge: Cambridge University Press. xiii, 387 p.
58. Olver, F.W.J. and National Institute of Standards and Technology (U.S.), *NIST handbook of mathematical functions*. 2010, Cambridge: Cambridge University Press. xv, 951 p.
59. Whittaker, E.T. and G.N. Watson, *A course of modern analysis : an introduction to the general theory of infinite processes and of analytic functions, with an account of the principal transcendental functions*. 4th ed. 1927, Cambridge: Cambridge University Press. 608 p.
60. Zhang, S. and J. Jin, *Computation of special functions*. 1996, New York; Chichester: Wiley. xxvi, 717 p.
61. Hairer, E., S.P. Nörsett, and G. Wanner, *Solving ordinary differential equations. I, Nonstiff problems*. 2nd rev. ed. Springer series in computational mathematics. 1993, Berlin ; London: Springer-Verlag. xv, 528 p.
62. Love, A.E.H., *A treatise on the mathematical theory of elasticity*. 2th ed. 1906, London: Cambridge university press
63. Sachkov, Y.L., *Maxwell strata in the Euler elastic problem.* Journal of Dynamical and Control Systems, 2008. **14**(2): p. 169-234.
64. Bateman Manuscript Project. California Institute of Technology., et al., *Higher transcendental functions, Vol 2*. 1953, New York ; London: McGraw-Hill.
65. Mutyalarao, M., D. Bharathi, and B.N. Rao, *On the uniqueness of large deflections of a uniform cantilever beam under a tip-concentrated rotational load.* International Journal of Non-Linear Mechanics, 2010. **45**(4): p. 433-441.
66. Batista, M. *Equilibrium Configurations of Cantilever Beam under Terminal Load*. 2013; Available from: http://www.maplesoft.com/applications/.